\documentclass[journal]{IEEEtran}
\pdfoutput=1
\usepackage{citesort}
\usepackage{amsmath,amssymb,amsfonts}
\usepackage{graphicx}
\usepackage{textcomp}
\usepackage{xcolor}
\usepackage{epstopdf}
\usepackage{subfigure}
\usepackage{verbatim}
\usepackage{algorithm}
\usepackage{algpseudocode}
\usepackage{setspace}
\usepackage{float}
\usepackage{lineno}
\usepackage{bm}

\newtheorem{lemma}{Lemma}
\begin{document}
\title{Power Optimization for Integrated Active and Passive Sensing in DFRC Systems}
\author{
Xingliang Lou,~\IEEEmembership{Student Member,~IEEE},
Wenchao Xia,~\IEEEmembership{Member,~IEEE},
Kai-Kit Wong,~\IEEEmembership{Fellow,~IEEE},
Haitao Zhao,~\IEEEmembership{Senior Member,~IEEE},
Tony Q. S. Quek,~\IEEEmembership{Fellow,~IEEE},
and Hongbo Zhu,~\IEEEmembership{Member,~IEEE}

\thanks{X. Lou, W. Xia, H. Zhao and H. Zhu are with the Jiangsu Key Laboratory of Wireless Communications, Nanjing University of Posts and Telecommunications, Nanjing 210003, P. R. China, (e-mail: 2022010109@njupt.edu.cn, xiawenchao@njupt.edu.cn, zhaoht@njupt.edu.cn, zhuhb@njupt.edu.cn).}
\thanks{K.-K. Wong is with the Department of Electronic and Electrical Engineering, University College London, London WC1E 6BT, U.K. He is also affiliated with Yonsei Frontier Laboratory, Yonsei University, Seoul, 03722, Korea, (e-mail : kai-kit.wong@ucl.ac.uk).}
\thanks{T. Q. S. Quek is with the Singapore University of Technology and Design, Singapore 487372, and also with the Yonsei Frontier lab, Yonsei University, south Korea (e-mail:tonyquek@sutd.edu.sg).}}
\maketitle
\begin{abstract}
  Most existing works on dual-function radar-communication (DFRC) systems mainly focus on active sensing, but ignore passive sensing.
  To leverage multi-static sensing capability, we explore integrated active and passive sensing (IAPS) in DFRC systems to remedy sensing performance. The multi-antenna  base station (BS) is responsible for communication and active sensing by transmitting signals to user equipments while detecting a target according to echo signals. In contrast, passive sensing is performed at the receive access points (RAPs). We consider both the cases where the capacity of  the backhaul links between the RAPs and BS is unlimited or limited and adopt different fusion strategies. Specifically, when the backhaul capacity is unlimited, the BS and RAPs transfer sensing signals they have received to the  central controller (CC) for signal fusion. The CC processes the signals and leverages the generalized likelihood ratio test detector to determine the present of a target. However, when the backhaul capacity is limited, each RAP, as well as the BS, makes decisions independently and sends its binary inference results to the CC for result fusion via voting aggregation. Then, aiming at maximize the target detection probability under communication quality of service constraints, two power optimization algorithms are proposed. Finally, numerical simulations demonstrate that the sensing performance in case of unlimited backhaul capacity is much better than that in case of limited backhaul capacity. Moreover, it implied that the proposed IAPS scheme outperforms only-passive and only-active sensing schemes, especially in unlimited capacity case.
\end{abstract}


\section{Introduction}

Communication networks are evolving from 5G to 6G in pursuit of a network that achieves global coverage, green intelligence, sensory interconnection, and synesthesia integration \cite{Zhang2019Wireless}. To achieve this vision, besides the communication ability, environmental perception ability is also required.
Electromagnetic wave has the ability to both sense environment and transmit data, but most existing works study and treat these two techniques independently, resulting in a conflict of wireless resources between sensing and communication systems.  In order to improve frequency spectrum and hardware efficiency, researchers are recently considering the function integration of wireless communication and radar sensing, which promotes the research on dual-function radar-communication (DFRC) systems \cite{Hassanien2016DualFunction,Liu2020JointRadar,Zhang2021Perceptive}.

The primary idea behind DFRC systems is to share infrastructure and resources between communication and sensing functions, thereby combining radar sensing with wireless communication \cite{Zhang2022Enabling,Wild2021Joint,Zhang2022Integration,Pin2021IntegratedSensing}.
Recently, DFRC systems have garnered significant attention from both the academia and industry and  there have been some  works on the design and performance analysis of DFRC systems,  e.g., \cite{Donnet2006CombiningMIMORadar,Liu2018MUMIMO,Liu2020JointTransmit,liu2022CRB,Bazzi2023Integrated,Huang2022IntegratedSensing,Guo2023JointCommunication,Li2023IntegratedSensing,An2023Fundamental}. Specifically, the feasibility of coexistence of multiple-input multiple-output (MIMO) radar and orthogonal frequency division multiplexing communication was demonstrated and transmit precoding was optimized to eliminate mutual interference between the sensing and communication signals in \cite{Donnet2006CombiningMIMORadar}. In \cite{Liu2018MUMIMO}, two operational options were proposed. The first option involved sensing and communication each occupying a subset of antennas and the radar signal was designed to fall into the null space of  downlink communication channel. The other option used a unified waveform for both radar and communication functions. In \cite{Liu2020JointTransmit}, the DFRC system transmitted the weighted sum of independent radar and communication symbols and formed multiple beams towards the target and the communication receivers. Different from \cite{Liu2020JointTransmit}, \cite{liu2022CRB} proposed to jointly optimize the DFRC waveform and precoding matrix towards realizing sensing and communication functions simultaneously and  the Cram$\acute{\textrm{e}}$r-Rao bound was used as a performance metric of target estimation. Then, \cite{Bazzi2023Integrated} proposed a novel DFRC waveform
design with peak-to-average power ratio control to maintain the similarity constraint of radar chirp signal. Later, the authors of \cite{Huang2022IntegratedSensing} leveraged intelligent reflecting surface (IRS) to improve both the performance of the DFRC and mobile edge computing. A IRS-aided DFRC system was proposed in \cite{Guo2023JointCommunication} for resource allocation in a coal mine scenario. Also, \cite{Li2023IntegratedSensing} proposed an SDP3 framework for the study of sensing-communication tradeoff with the particular application of human motion recognition on indoor settings. Furthermore, \cite{An2023Fundamental} described the fundamental performance tradeoff between the detection probability for target sensing and the communication achievable rate and derived closed-form expressions for the detection probability in DFRC systems.
However, only signals of mono-static sensing transceiver was used in \cite{Donnet2006CombiningMIMORadar,Liu2018MUMIMO,Liu2020JointTransmit,liu2022CRB,Huang2022IntegratedSensing,Guo2023JointCommunication,Li2023IntegratedSensing,Bazzi2023Integrated,An2023Fundamental}. To improve sensing accuracy,  \cite{Ni2021Uplink,chowdary2023hybrid,xu2023integrated} attempted to make use of multiple signals for sensing. In particular, \cite{Ni2021Uplink} proposed an uplink sensing scheme which jointly processed all measurements from the spatial, temporal, and frequency domains for perceptive mobile networks with asynchronous transceivers. In \cite{chowdary2023hybrid}, a base station (BS) working as a mono-static radar receiver was used to estimate angles-of-arrival of targets based on  its downlink echo signals and uplink reflected signals from the users. To address the self-interference problem of echo signal  caused by the concurrent information transmission, the authors of \cite{xu2023integrated} proposed to select one BS as a receiver to receive the echo signals while other BSs act as transmitters.

Different from the above works which are based on mono-static sensing, recently multi-static sensing has attracted growing interests and is expected to bring various advantages over the conventional mono-static sensing. Multi-static sensing can not only reduce the mono-static sensing uncertainties caused by noise or incompleteness due to wireless fading and interference \cite{Gurbuz2019RadarBased}, but also provide better sensing coverage and capture richer sensing information \cite{godrich2010target}.  There are only a few works on multi-static sensing in DFRC systems.  In \cite{2022PowerAllocation}, a centralized federated processing method was proposed for the information received at multiple receive access points (RAPs) in a cell-free massive multiple-input multiple-output (MIMO) system. Also, the authors of \cite{yajnanarayana2022multistatic} attempted to improve sensing accuracy of human-sized objects by increasing the number of receive devices under indoor cellular deployment. Then, the authors of \cite{Huang2022Coordinated} utilized a set of distributed transmitters which sent individual information for multiple sensing receivers to estimate the target. Assuming time synchronization among BSs in \cite{cheng2022coordinated}, each BS can exploit reflected signals of itself and other BSs for joint detection. A method was proposed in \cite{li2022multi} to fuse the outputs of multiple dual-function radars to achieve higher sensing performance, but  the communication and sensing signals were sent in different time slots. Despite the above progress, these previous works still suffer from many limitations. First of all, although multi-static sensing provides enhanced sensing capability, we note that most existing works focus on active sensing in DFRC systems and ignore the potential performance gain from passive sensing. In multi-static sensing, we refer to the sensing operation based on echo signals as active sensing, whereas the sensing operation based on the received signals from other transmitters (such as BSs and RAPs) as passive sensing. Secondly, challenges are still present in terms of wireless resource allocation to properly balance the performance between sensing and communication. Finally,  direct transmission of multi-static sensing signals to a central controller (CC)  for centralized processing leads to high communication overhead. However, the channel links to the CC are often capacity-limited.

Motivated by the analysis above, we aim to improve sensing performance  by  integrated active and passive sensing (IAPS) in DFRC systems without sacrificing communication performance.  Specifically, the BS is responsible for communication as well as active sensing by transmitting signals to user equipments (UEs) while simultaneously detecting targets according to the echo signals. RAPs do not transmit signals to UEs but they can provide reflected signals.  To leverage the multi-static sensing capability and address the uncertainty in sensing due to fading and interference characteristics of wireless communication, we explore the performance gain of passive sensing performed at RAPs based on the reflected signals they have received. Considering the fact that the capacity of backhaul links between the CC and RAPs are usually limited, it is hard for each RAP to send its observation (i.e., reflected signals they have received) directly to the CC. This practical consideration poses a challenge on the integration of active and passive sensing.

In this work, we consider both the cases with unlimited and limited backhaul capacity and propose different fusion strategies. The contributions of this work are summarized as follows:
\begin{enumerate}
  \item
  We consider a DFRC system where a BS communicates with UEs  and  senses a target simultaneously. Multiple RAPs are connected to the CC via backhaul links. Moreover,  we assume that the RAPs and the BS are perfectly synchronized.
  In addition to active sensing signals received at the BS,  passive sensing signals received at the RAPs are also exploited and then the IAPS scheme is proposed to improve sensing performance under communication quality-of-service (QoS) constraints. Furthermore, we consider two cases, i.e., one with unlimited backhaul capacity and the other  with limited backhaul capacity, and propose power optimization algorithms, respectively.
  \item
  In the case where the backhaul capacity is unlimited, in order to improve the sensing performance, the CC collects the sensing signals received from the BS and RAPs for signal fusion. All processing is done centrally in the CC. Then, the CC processes the received signals and uses the generalized likelihood ratio test (GLRT) detector to determine if a target is present. Moreover, we propose a power allocation algorithm to maximize the detection probability for improving sensing performance.
  \item
  In the case where the backhaul capacity is limited, the  BS and RAPs make decisions based on their observation independently and send binary inference results to the CC for result fusion. A whitening filter  is adopted to eliminate the direct path interference \cite{Liu2018MIMORadar}. Upon receiving active and passive sensing binary inference results, the CC performs voting aggregation to determine whether the target exists.  We convert the probability of error minimization into a maximization problem of joint detection probability  and propose a heuristic power optimization algorithm.
  \item
  Finally, numerical simulations in the cases with unlimited and limited backhaul capacity are conducted. Numerical results demonstrate that  the sensing performance in the case of unlimited backhaul capacity is much better than that in the case of limited backhaul capacity. It is also observed that the performance of the proposed IAPS scheme is better than that of  only-active and only-passive sensing schemes especially when the backhaul capacity is unlimited. Besides, we find that the overall performance can be improved by increasing the number of RAPs.

\end{enumerate}

The structure of this paper is given as follows. Firstly, Section II  presents the system model. Then, Sections III and IV formulate the sensing performance optimization problems with communication QoS constraints in the cases where the backhaul capacity is unlimited and limited, respectively, and then propose corresponding power allocation algorithms. Numerical results are provided and discussed in Section V. Finally, Section VI concludes this paper.

\emph{Notation}: Matrices and vectors are represented by bold uppercase and lowercase symbols, respectively. $\mathbb{E}(\cdot)$ denotes the expectation operator. $\mathrm{diag}(\cdot)$ and $\mathrm{blkdiag}(\cdot)$  stand for the construction of a diagonal matrix and the construction of a block diagonal matrix, respectively. $\|\cdot\|_1$ and $\|\cdot\|_2$ represent $l_1$ and $l_2$ norm, respectively. $\operatorname{tr}(\cdot)$ and $\operatorname{vec}(\cdot)$
denote the trace and the vectorization operations, respectively. $\mathfrak{R}(\cdot)$ and $\mathfrak{I}(\cdot)$ denote the real and imaginary parts of the argument.  $(\cdot)^T $ and $(\cdot)^H$ stand for transpose and Hermitian transpose of the matrices, respectively. $\mathbf{I}$ is the identity matrix. $ \lceil \cdot \rceil$ denotes the ceiling function.

\section{System Model}
\subsection{System Setting}
\begin{figure}[htbp]
	\centerline{\includegraphics[width=0.53\textwidth]{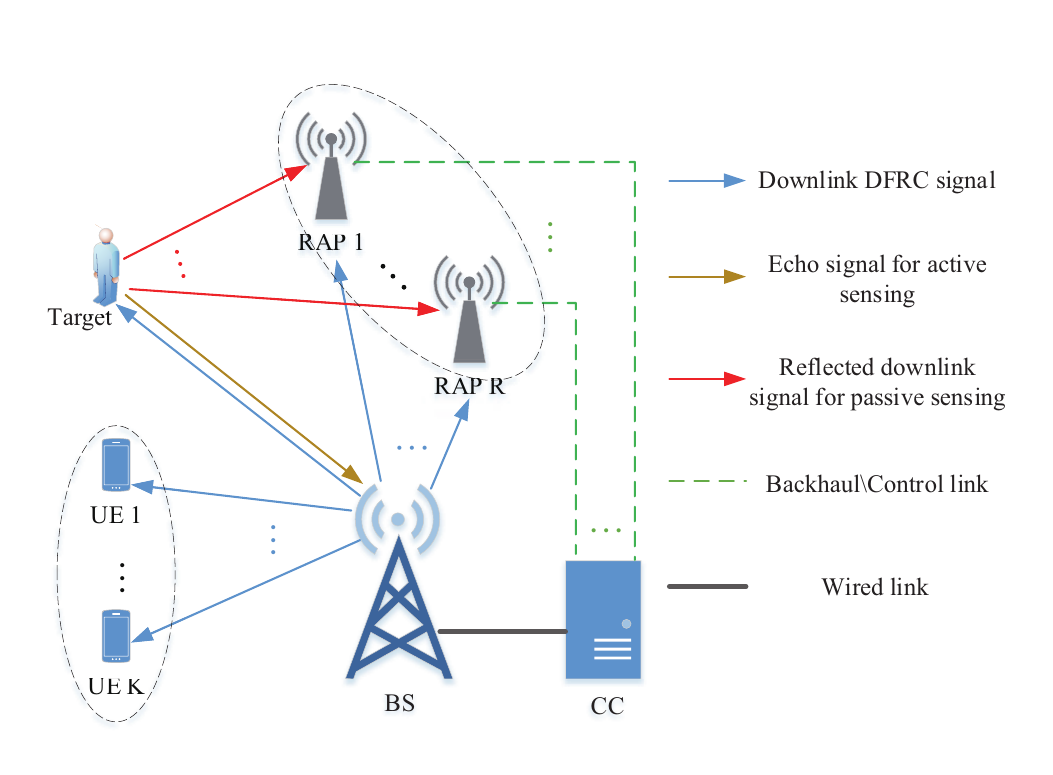}}
	\caption{ Illustration of IAPS in DFRC System comprising $K$ UEs, $R$ RAPs, a DFRC BS and a CC, where the RAPs and the BS are fully synchronized controlled through the CC. Specifically, the BS transmit DFRC signals to serve the $K$ UEs and sense a desired target, while $R$ RAPs and the BS receive echo signal.
}
	\label{scenario}
\end{figure}

We consider a downlink DFRC system, as depicted in Fig. \ref{scenario}, where a BS equipped with $M$ transmit antennas and $N_{0}$ receive antennas is responsible for serving $K$ single-antenna UEs and detecting a single target simultaneously. There are also $R$ RAPs each with $N_{1}$ receive antennas, which can be used to receive the reflected signals for sensing. Here, we refer to the sensing operation at the BS and RAPs as active and passive sensing, respectively.  Note that $K \le  M$ is required to ensure the feasibility of downlink multi-user MIMO communications and $M < N_{0}$ is typically required to avoid the loss of information about the target \cite{liu2022CRB}. Besides,  a CC is introduced to process both the active and passive sensing signals. For notational convenience, we denote the set of UEs and the set of RAPs by $\mathcal{K}=\{1,2,\ldots,K\}$ and $\mathcal{R}=\{1,2,\ldots,R\}$, respectively. The BS is indexed using 0 and we denote the set of the BS and RAPs by $\mathcal{R}^{\prime}=\mathcal{R}\bigcup\{0\}$, Besides, the target is indexed using 0 and the set of UEs and target are denoted by $\mathcal{K}^{\prime}=\mathcal{K}\bigcup\{0\}$.

Define $\mathbf{s}_0\in \mathbb{C}^{L\times 1}$ as the dedicated sensing symbol vector and $\mathbf{s}_k\in \mathbb{C}^{L\times 1}$ as the communication symbol vector of  the $k$-th UE, respectively, with $L$ being the length of the communication time slots. It is assumed that the sensing and communication symbols are independent with $\frac{1}{L}\mathbb{E}(\mathbf{S} \mathbf{S} ^H) \approx \mathbf{I}_{K+1}$, where $\mathbf{S}=[\mathbf{s}_0,\mathbf{s}_1,...,\mathbf{s}_K]^T$, which holds for white Gaussian signals and sufficiently large $L$ value.  Then, the DFRC signal matrix is given as
$\mathbf{X}=\mathbf{W}\mathbf{S}$, where
\begin{equation}
\mathbf{W}=[ \sqrt{p_0}\mathbf{\tilde{w}}_0,\sqrt{p_1}\mathbf{\tilde{w}}_1,...,\sqrt{p_K}\mathbf{\tilde{w}}_K] \in \mathbb{C}^{{M}\times  ( K+1  )},
\end{equation}
where $p_k$ and $\mathbf{\tilde{w}}_k$ are the transmit power and the  normalized precoding vector for target and the $k$-th UE for all $k\in\mathcal{K}^{\prime}$, respectively, with $\|\mathbf{\tilde{w}}_k\|_2=1$.

\subsection{Communication Model}
The received signal in the $l$-th communication symbol at the $k$-th UE is given as
\begin{equation}
y_k[l] =\mathbf{h}_k^H\sum_{k'=1}^{K}\sqrt{p_{k'}}\mathbf{\tilde{w}}_{k'}\mathbf{s}_{k'} [l] +\sqrt{p_0}\mathbf{h} _k^H\mathbf{\tilde{w}} _0\mathbf{s}_0[l] +\mathbf{n}_k[l] ,
\end{equation}
where $\mathbf{n}_k[l] \sim \mathcal{CN}(0,\sigma_{\rm n} ^2)$ denotes the additive white gaussian noise (AWGN) with zero mean and variance $\sigma_{\rm n}^2$ and $\mathbf{h}_k$ denotes  the channel between the $k$-th UE and the BS, which is assumed to be flat Rayleigh fading and statistically independent of each other.

Then, the signal to interference plus noise ratio (SINR) of the $k$-th UE is given by
\begin{equation}\label{sinr}
\gamma _k=\frac{p_k|\mathbf{h} _k^H\mathbf{\tilde{w}} _k |^2 }{\sum_{k'=1,k'\ne k}^{K} p_{k'}| \mathbf{h} _k^H\mathbf{\tilde{w}} _{k'}|^2+p_0| \mathbf{h} _k^H\mathbf{\tilde{w}} _0|^2 +\sigma_{\rm n} ^2   }.
\end{equation}\par

\subsection{Sensing Model}

The BS can sense the target through the echo signal, which is given as
\begin{equation}
\mathbf{z} _0[l]=\alpha _0\mathbf{b}_0 ( \theta ) \mathbf{a} ^H ( \theta ) \mathbf{X}[l]+\mathbf{{n}}^\prime _0[l]\in \mathbb{C}^{N_0\times 1},
\end{equation}
where $\mathbf{{n}}^\prime _0[l]\sim\mathcal{C N}(\mathbf{0}, \sigma_{\rm n}^{2}\mathbf{I}_{N_0})$ denotes the AWGN vector, $\mathbf{X}[l]$ represents the $l$-th column of $\mathbf{X}$, $\alpha_{0}$ is the combined sensing channel gain that includes the path-loss through target and the radar cross section (RCS) of the target \cite{Zhao2022JointTransmit} which follows Swerling-I model \cite{1987Principles}, in which the target is static or has a very slow mobility, and $\theta$ is the azimuth angle of target relative to the antenna array at the BS. The transmit and receive  steering vectors of the BS are denoted by
\begin{equation}
\mathbf{a}(\cdot )=\big[1, e^{j 2 \pi \delta \sin (\cdot)}, \ldots, e^{j 2 \pi(M-1) \delta \sin (\cdot)}\big]^{T} \in \mathbb{C}^{M\times 1},
\end{equation}
\begin{equation}
\mathbf{b}_0(\cdot )\!=\!\big[1, e^{j 2 \pi \delta \sin (\cdot)}, \ldots, e^{j 2 \pi(N_0-1) \delta \sin (\cdot)} \big]^{T} \!\in\! \mathbb{C}^{N_0\times 1},
\end{equation}
respectively, where $\delta$ is the spacing between adjacent antennas normalized by wavelength. \par

When a target is present, the reflected sensing signal received at the $r$-th RAP for passive sensing is given as
\begin{equation}\label{sensing signal}
\mathbf{z} _r[l]=\alpha _r\mathbf{b}_1 ( \varphi _r ) \mathbf{a} ^H(\theta) \mathbf{X}[l]+\mathbf{G}_r \mathbf{X}[l]+\mathbf{{n}}^\prime _r[l]\in \mathbb{C}^{N_1\times 1},
\end{equation}
where $\mathbf{n}^{\prime}_r[l]\sim\mathcal{C N}(\mathbf{0}, \sigma_{\rm n}^{2}\mathbf{I}_{N_1})$ denotes the AWGN vector, $\alpha_{r}$ is the combined sensing channel gain which follows the model similar to $\alpha_{0}$ for the convenience of analysis, given as
\begin{equation}
\alpha_{r} \sim \mathcal{C N}(0, \sigma_{\rm rcs}^{2}),r\in \mathcal{R}^{\prime},
\end{equation}
$\varphi _r$ is the azimuth angle of target relative to the $r$-th RAP, $\mathbf{G}_r \in \mathbb{C}^{N_1\times M}$  represents the target-free channel between the BS and the  $r$-th RAP and is assumed to follow complex Gaussian distribution, and $\mathbf{b}_1(\cdot )$ is the steering vector of the RAPs, given as
\begin{equation}
\mathbf{b}_1(\cdot )\!=\!\big[1, e^{j 2 \pi \delta \sin (\cdot)}, \ldots, e^{j 2 \pi(N_1-1) \delta \sin (\cdot)}\big]^{T} \!\in\! \mathbb{C}^{N_1\times 1}.
\end{equation} \par

\subsection{Transmit Precoding Vectors}
The transmit precoding  is designed  based on the regularized zero-forcing (RZF) scheme, i.e., $\mathbf{\tilde{w}}_k=\bar{\mathbf{w} }_k / \| \bar{\mathbf{w} }_k \|_2 $ with
\begin{equation}
\bar{\mathbf{w} }_k= ( \mathbf{H}\mathbf{H}^H+\lambda \mathbf{I}_{M}) ^{-1}\mathbf{h} _k,
\end{equation}
where
$\mathbf{H} = [ \mathbf{h}_1,\mathbf{h}_2,...,\mathbf{h}_K   ] \in \mathbb{C}^{M\times K}$
 and $\lambda$ is the regularization parameter. \par

In order to eliminate the interference caused by sensing symbols to the UEs, we employ the zero-forcing radar (ZFR) precoder
$\mathbf{\tilde{w}}_0= \bar{\mathbf{w} }_0/ \| \bar{\mathbf{w} }_0 \|_2 $ \cite{Buzzi2019MassiveMIMO},
where
\begin{equation}\label{w0}
\bar{\mathbf{w} }_0=(\mathbf{I}_M-\mathbf{H} \mathbf{H}^H) ^{-1}\mathbf{a} (\theta).
\end{equation}\par

The BS and RAPs are connected through the backhaul channel using wired or wireless links. In this work, we conduct analysis in two cases, i.e., unlimited and limited backhaul capacity. Specifically, when the backhaul capacity is unlimited (such as fiber links), both the BS and  RAPs directly send the received sensing signals to the CC for joint processing, which will be introduced in Section III. However, when the backhaul capacity is limited (such as wireless links), it is impractical for the RAPs to send the sensing signal directly due to the large amount of sensing signals. As an alternative solution,  each RAP first makes decisions independently and then send binary inference results to the CC for voting aggregation, such that only a few bits are needed to exchange. More details about the latter solution will be given in Section IV.\par

\section{DFRC System With Unlimited Backhaul Capacity}
In this section, we assume the backhaul capacity is unlimited. Thus,  the BS and  RAPs can directly send the received sensing signals to the CC for joint processing.
\subsection{Active and Passive Sensing Signal Fusion}
With the unlimited backhaul capacity, it is reasonable to assume that $\mathbf{X}$ and $\mathbf{G}=[\mathbf{G}_1,...,\mathbf{G}_R]$  are known at the CC, which can be estimated using the method proposed in \cite{Li2017JointTransmit}. Therefore, the target-free portion of the received sensing signal at the RAPs can be mitigated perfectly. Then, $\mathbf{z} _r$ in  \eqref{sensing signal} can be rewritten as
\begin{align}
\mathbf{z} _r[l]=\alpha _r\mathbf{A} _r[l]+\mathbf{{n}}'_r[l],
\end{align}
where
$
\mathbf{A} _r[l]=\mathbf{b}_1  ( \varphi _r  ) \mathbf{a} ^H ( \theta   ) \mathbf{X}[l]
$
is the known part at the CC. For subsequent calculations, we also define $\mathbf{B}_0=\mathbf{b}_0  ( \theta  ) \mathbf{a} ^H ( \theta   )$ and
$\mathbf{B}_r=\mathbf{b}_1  ( \varphi _r  ) \mathbf{a} ^H ( \theta   )$.\par

We collect the received  sensing signals at the BS and RAPs into a vector $\mathbf{z}[l]=\big [ \mathbf{z}_0^T[l]\dots \mathbf{z}_R^T[l] \big ]^T$, and the overall sensing signal is expressed as
\begin{equation}
\mathbf{z}[l]=\mathbf{A}[l]\boldsymbol{\alpha}+\mathbf{{n}'}[l] \in \mathbb{C}^{(N_0+RN_1)\times 1},
\end{equation}
where
$
\boldsymbol{\alpha} = [ \alpha _0,\alpha _1,...,\alpha _R  ]^T
$ is a collection of the  unknown sensing channel coefficients, $\mathbf{n}^{\prime}[l]=\big[ \mathbf{n}^{\prime T}_0 [l]\dots \mathbf{n}^{\prime T}_R[l]  \big]^T $ is the concatenated noise, and $\mathbf{A}[l]$ is the known part and given as
\begin{equation}
\mathbf{A}[l]\!\!=\!\!\mathrm{blkdiag}\big(\!\mathbf{A}_0[l],\mathbf{A}_1[l],...,\mathbf{A}_R[l]\!\big)\!\!\in\! \mathbb{C}^{(\!N_0\!+\!RN_1\!)\!\times\! (\!R\!+\!1\!)}.
\end{equation}

\subsection{GLRT Detector}
Here, the received signals in $L$ communication time slots are used for detection. We define the vectors $\mathbf{z} _L\in \mathbb{C}^{(N_0+RN_1)L\times 1}$, $\mathbf{n}' _L\in \mathbb{C}^{(N_0+RN_1)L\times 1}$, and $\mathbf{c} '_L\in \mathbb{C}^{(N_0+RN_1)L\times 1}$, which are constructed by concatenating $\mathbf{z}[l]$, $\mathbf{n}'[l]$, and $\mathbf{A}[l]\bm{\alpha}$, respectively, i.e.,
\begin{equation}
\mathbf{z}_L= \big[ \mathbf{z}^T[1]\dots \mathbf{z}^T[L]  \big]^T,
\end{equation}
\begin{equation}
\mathbf{n}' _L= \big[ \mathbf{n}'^T[1]\dots \mathbf{n}'^T[L]  \big]^T.
\end{equation}
and
\begin{equation}
\mathbf{c} '_L=\Big [\big[\mathbf{A}[1]\boldsymbol{\alpha}\big]^T\dots \big[\mathbf{A}[L]\boldsymbol{\alpha}\big]^T \Big ]^T,
\end{equation}\par
The binary hypothesis used in the GLRT detector is written as
\begin{equation}
\begin{cases}
\mathcal{H}_{0}: \mathbf{z}_{L}=\mathbf{n}_{L}^{\prime}, \\
\mathcal{H}_{1}: \mathbf{z}_{L}=\mathbf{A}'_{L}+\mathbf{n}_{L}^{\prime}.
\end{cases}
\end{equation}
Then the corresponding GLRT detector is given by
\begin{equation}
\Lambda =\frac{\max _{\boldsymbol{\alpha}}f(\mathbf{z}_L|\boldsymbol{\alpha},\mathcal{H} _1 ) }{f(\mathbf{z}_L|\mathcal{H} _0 )}
\mathop{\gtrless}\limits_{\mathcal{H} _0}^{\mathcal{H} _1}\xi,
\end{equation}
where $\xi $ is the threshold of the GLRT detector, which is selected to achieve a desired false alarm probability $P_{\rm FA}$ under Neyman-Pearson criterion \cite{1993Fundamentals}.
The joint probability density function (PDF) of the overall received sensing signal in cases $\mathcal{H}_1$ and $\mathcal{H}_0$ is computed as \cite{He2010MIMORadar}
\begin{align}
f(\mathbf{z}_L|&\boldsymbol{\alpha},\mathcal{H} _1 )=\frac{1}{(\pi \sigma _{\rm n}^2)^{(R+1)L}}
\cdot \nonumber \\
&\exp \Bigg(- \sum_{l=1}^{L}\frac{\big(\mathbf{z}[l]-\mathbf{A}[l] \boldsymbol{\alpha}\big)^H\big(\mathbf{z}[l]-\mathbf{A}[l] \boldsymbol{\alpha}\big)}{\sigma _{\rm n}^{2}} \Bigg),
\end{align}
and
\begin{equation}
f(\mathbf{z}_L|\mathcal{H} _0 )=\frac{1}{(\pi \sigma _{\rm n}^2)^{(R+1)L}}
\cdot \exp\Bigg( - \sum_{l=1}^{L}\frac{\mathbf{z}[l]^H\mathbf{z}[l]}{\sigma _{\rm n}^{2}}\Bigg),
\end{equation}
respectively. We note that maximizing $f(\mathbf{z}_L|\boldsymbol{\alpha},\mathcal{H} _1 )$ with respect to $\boldsymbol{\alpha}$ is equivalent to
\begin{equation}
\min_{\boldsymbol{\alpha}}\sum_{l=1}^{L}\frac{\big(\mathbf{z}[l]-\mathbf{A} [l]\boldsymbol{\alpha}\big)^H\big(\mathbf{z}[l]-\mathbf{A} [l]\boldsymbol{\alpha}\big)}{\sigma_{\rm n}^2},
\end{equation}
which in turn is equivalent to
\begin{equation}
\max_{\boldsymbol{\alpha}}2\mathfrak{R}\Bigg(\! \boldsymbol{\alpha}^H\sum_{l=1}^{L}\!\frac{\mathbf{A}^H [l]\mathbf{z}[l]}{\sigma _{\rm n}^2} \!\Bigg)\!-\!\boldsymbol{\alpha}^H\Bigg(\!\sum_{l=1}^{L}\!\frac{\mathbf{A}^H [l]\mathbf{A}[l]}{\sigma _{\rm n}^2}\!\Bigg)\boldsymbol{\alpha}.
\end{equation}
Thus, $\boldsymbol{\alpha}$ can be estimated as
\begin{equation}
    \hat{\boldsymbol{\alpha}} =\Bigg(\sum_{l=1}^{L}\mathbf{A}^H[l]\mathbf{A}[l]\Bigg)^{-1}\Bigg(\sum_{l=1}^{L}\mathbf{A}^H[l]\mathbf{z}[l] \Bigg).
\end{equation}
Inserting the estimated $\hat{\boldsymbol{\alpha}}$ into the GLRT detector, we obtain the test statistic as
\begin{align}
&\mathrm{ln} (\Lambda )=\nonumber \\
&\frac{1}{\sigma _{\rm n}^2}\!\Bigg(\!\sum_{l=1}^{L}\!\mathbf{z}^H[l]\!\mathbf{A}[l]\!\Bigg)\!
\Bigg(\!\sum_{l=1}^{L}\!\mathbf{A}^H[l]\!\mathbf{A}[l]\!\Bigg)^{-1}\!\Bigg(\!\sum_{l=1}^{L}\!\mathbf{A}^H[l]\mathbf{z}[l]\!\Bigg).
\end{align}

According to \cite{1993Fundamentals}, the asymptotic distribution of $\mathrm{ln} (\Lambda )$  can be expressed as
\begin{equation}
    \mathrm{ln} (\Lambda ) \sim\left\{\begin{array}{l}
\mathcal{H}_{1}: \mathcal{X}_{2}^{2}(\rho), \\
\mathcal{H}_{0}: \mathcal{X}_{2}^{2},
\end{array}\right.
\end{equation}
where $\mathcal{X}_{2}^{2}$ and $\mathcal{X}_{2}^{2}(\rho)$ are central and non-central chi-squared distributions with two Degrees of Freedom (DoFs), respectively, and $\rho$ is the non-central parameter. When the GLRT is used, the threshold $\xi$ can be expressed as
\begin{equation}
    \xi=\mathfrak{F}_{\mathcal{X}_{2}^{2}}^{-1}(1-P_{\rm FA}),
\end{equation}
and the detection probability $P_{\rm D}$ is given as \cite{Bekkerman2006TargetDetection}
\begin{equation}\label{p_d}
P_{\rm D}=1-\mathfrak{F}_{\mathcal{X}_{2}^{2}(\rho)}(\xi),
\end{equation}
where $\mathfrak{F}_{\mathcal{X}_{2}^{2}(\rho)}$ is the non-central chi-square Cumulative Distribution Function (CDF) with two DoFs.

Because different RCSs are assumed to be independent and have zero mean, we have
\begin{align}
& \mathbb{E}\big(\mathbf{A}[l]\boldsymbol{\alpha}\boldsymbol{\alpha}^H\mathbf{A}^H[l]\big)\nonumber\\
&=\mathbb{E} \Big(\mathrm{blkdiag}\big( | \alpha _0  |^2\mathbf{A}_0[l]\mathbf{A}_0[l]^H,..., | \alpha _R  |^2\mathbf{A}_R[l]\mathbf{A}_R[l]^H \big)\Big)\nonumber\\
&=\sigma_{\rm rcs}^{2}\mathrm{blkdiag}(\mathbf{B}_0\mathbf{\hat{W}}\mathbf{B}^H_0,...,\mathbf{B}_R\mathbf{\hat{W}}\mathbf{B}^
H_R),
\end{align}
where
$\mathbf{\hat{W}}=\mathbf{W}\mathbf{W}^H$.
Then, the non-centrality parameter $\rho$ is given as \cite{1993Fundamentals}
\begin{align}
\rho&=\sum_{l=1}^{L}\operatorname{tr}\Big(\mathbb{E} \big(\mathbf{A}[l]\boldsymbol{\alpha}\boldsymbol{\alpha}^H\mathbf{A}^H[l]\big)\big(\mathbb{E}(\|\mathbf{{n}'}[l]\|_2^2) \big)^{-1}\Big )\nonumber\\
&=\frac{L\sigma_{\rm rcs}^{2}}{(N_0+RN_1)\sigma _{\rm n}^2} \operatorname{tr}\big(\mathrm{blkdiag}(\mathbf{B}_0\mathbf{\hat{W}}\mathbf{B}^H_0,...,\mathbf{B}_R\mathbf{\hat{W}}
\mathbf{B}^H_R) \big)\nonumber\\
&=\frac{L\sigma_{\rm rcs}^{2}}{(N_0+RN_1)\sigma _{\rm n}^2}\sum_{r=0}^{R}\operatorname{tr}( \mathbf{B}_{r} \hat{\mathbf{W}} \mathbf{B}_{r}^{H}).
\end{align}
\par

\subsection{Problem Formulation}
The goal is to maximize the detection performance under SINR constraints of the UEs and a power constraint of the BS, which is formulated as
\begin{subequations}
\begin{align}
\mathcal{P}_0: \ &\underset{\boldsymbol{\mathbf{p}\ge 0 }}{\operatorname{max}} \ P_{\rm D} \nonumber \\
 \text {s.t.}\ &\gamma_{k} \geq \Gamma, k\in \mathcal{K}, \label{qos constraint} \\
& ||\mathbf{p}||_1 \leq P_{\max} \label{power constraint},
\end{align}
\end{subequations}
where $\mathbf{p } = [ p_0,p_1,...,p_K  ]^T$, $\Gamma$  is the minimum SINR threshold for the UEs, and $P_{\max}$ is the power budget of the BS. It is observed that $P_{\rm D}$ is a monotonically increasing function with respect to $\rho$ \cite{1993Fundamentals} and $\frac{L\sigma_{\rm rcs}^{2}}{(N_0+RN_1)\sigma _{\rm n}^2}$ is a constant. Therefore, problem $\mathcal{P}_0$ can be equivalently formulated as

\begin{align*}
\mathcal{P}_1:\ &\underset{\boldsymbol{\mathbf{p}\ge 0 }}{\operatorname{max}} \  \sum_{r=0}^{R}\operatorname{tr}( \mathbf{B}_{r} \hat{\mathbf{W}} \mathbf{B}_{r}^{H})  \\
 \text {s.t.} \ &\eqref{qos constraint} \ \text{and} \ \eqref{power constraint}.
\end{align*}\par
According to the properties of matrix trace, we have
\begin{align}
\operatorname{tr}(\mathbf{B}_{r} \hat{\mathbf{W}} \mathbf{B}_{r}^{H})& =\operatorname{tr}(\hat{\mathbf{W}} \mathbf{B}_{r}^{H}\mathbf{B}_{r} ) \nonumber\\
&= \operatorname{tr}\big(\mathrm{diag}(\mathbf{p })\mathbf{\tilde{W}}^H \mathbf{B}_{r}^{H}\mathbf{B}_{r}\mathbf{\tilde{W}} \big),
\end{align}
where
$
\mathbf{\tilde{W}}=[\mathbf{\tilde{w}}_0,\mathbf{\tilde{w}}_1,...,\mathbf{\tilde{w}}_K ].
$
Since $\mathbf{\tilde{W}}^H \mathbf{B}_{r}^{H}\mathbf{B}_{r}\mathbf{\tilde{W}}$ is a Hermitian symmetric matrix where the elements on the diagonal must be real, i.e.,
$
\operatorname{tr}\big(\mathfrak{I}(\mathbf{\tilde{W}}^H\mathbf{B}_{r}^{H}\mathbf{B}_{r}\mathbf{\tilde{W}})\big)=0,
\forall \mathbf{p}$.
The QoS constraints can be rewritten in the form of second-order cone as \cite{2022PowerAllocation}
\begin{align}\label{The first set rewritten}
 \big\| [\varrho_{k, 0}\sqrt{p_{0}}, \ \varrho_{k, 1}\sqrt{p_{1}}, \ \dots, \ 0, \ \dots,\ \varrho_{k, K}\sqrt{p_{K}},\ \sigma_{\rm n} ] \big\|_2 \nonumber\\
\leq \cfrac{\sqrt{p_{k}} \varrho_{k, k}}{\sqrt{\Gamma}},k=1, \ldots, K,
\end{align}
where
$\varrho_{k, j}=|\mathbf{h}_{k}^{H} \mathbf{\tilde{w}}_{j}|$. Now, problem $\mathcal{P}_1$ is reformulated as
\begin{align*}
\mathcal{P}_2: \ &\underset{\boldsymbol{\mathbf{p}\ge 0 }}{\operatorname{min}}-\sum_{r=0}^{R}\operatorname{tr}\big(\mathrm{diag}(\mathbf{p })\mathfrak{R} (\mathbf{\tilde{W}}^H \mathbf{B}_{r}^{H}\mathbf{B}_{r}\mathbf{\tilde{W}}) \big)\\
 \text {s.t.} \ &\eqref{power constraint} \ \text{and} \ \eqref{The first set rewritten}.
\end{align*}
Now, $\mathcal{P}_2$ becomes a standard Semidefinite Program (SDP) which can be solved using optimization tools, such as CVX \cite{2004Convex}.\par

\section{ DFRC System With Limited Backhaul Capacity }
When the backhaul capacity is limited, the RAPs cannot directly send the reflected signals they have received. Instead, they independently make decisions based on their own observation and then send binary inference results to the CC for result fusion via voting aggregation. Moreover, since the BS and RAPs are network infrastructure and belong to network operators, it is reasonable to assume that the BS and RAPs share a public set of random symbols. After the CC selects the sensing and communication symbols from the known symbol set, it sends the indices of the selected symbols to the RAPs. Then, the RAPs determine a matching filter based on the indices.

\subsection{GLRT Detector}
Since the backhaul capacity is limited and the RAPs have to make decisions locally, the interference to the sensing signal at the RAPs from the BS via the target-free channel cannot be ignored. Then, the binary hypothesis is  described as
\begin{equation}\label{GLRT2}
\begin{cases}
 \mathcal{H}_0:\mathbf{z}_r[l]=\mathbf{G}_r\mathbf{X}[l]+\mathbf{n}'_r[l],\\
  \mathcal{H}_1:\mathbf{z}_r[l]=\alpha_r\mathbf{B}_r\mathbf{X}[l]+\mathbf{G}_r\mathbf{X}[l]+\mathbf{n}'_r[l].
\end{cases}
\end{equation}\par
For simplicity, we assume that the RAPs have accurately estimated the interference-plus-noise covariance matrix \cite{Liu2018MIMORadar}. We use the GLRT detector to solve the unknown parameters $\alpha_r$, $\varphi _r$ and $\theta$. In order to consider the sufficient statistic of the received signal, a matching filter (i.e. $\mathbf{S}$ ) is adopted \cite{Khawar2015TargetDetection}, which is given as,

\begin{align}
\tilde{\mathbf{Z}}_r &=\frac{1}{\sqrt{L}} \sum_{l=1}^{L} \mathbf{z}_{r}[l] \mathbf{S}^H[l]\nonumber\\
&=\alpha_r \sqrt{L } \mathbf{B}_r\mathbf{W} +\frac{1}{\sqrt{L}} (\mathbf{G}_r\mathbf{X}+\mathbf{n}'_r)\mathbf{S}^H.
\end{align}

Define $\tilde{\mathbf{z}}$ as the vectorization of $\tilde{\mathbf{Z}}$, which is given as
\begin{align}
\tilde{\mathbf{z}}_r&=\mathrm{vec} (\tilde{\mathbf{Z}}_r)\nonumber\\
&=\alpha_r \sqrt{L } \mathrm{vec} (\mathbf{B}_r\mathbf{W})+\boldsymbol{\varepsilon}_r,
\end{align}
where $\boldsymbol{\varepsilon}_r=\frac{1}{\sqrt{L}} \mathrm{vec} \big((\mathbf{G}_r\mathbf{X}+\mathbf{n}'_r)\mathbf{S}^H\big)$ follows zero-mean, complex Gaussian distribution and has the following block covariance matrix \cite{Liu2018MIMORadar}:
\begin{equation}
    \mathbf{C}_r\!=\!\left[\!\!\begin{array}{ccc}
\mathbf{Q}_r\!+\!\sigma_{\rm n}^{2} \mathbf{I}_{N_1} & & \mathbf{0} \\
&\!\cdots\!\\
\mathbf{0} &  & \mathbf{Q}_r\!+\!\sigma_{\rm n}^{2} \mathbf{I}_{N_1}
\end{array}\!\!\right]\!\in\! \mathbb{C}^{N_1M\!\times\! N_1M},
\end{equation}
where $\mathbf{Q}_r=\mathbf{G}_r \hat{\mathbf{W}}  \mathbf{G}_r^{H}$.

Before using the GLRT detector,  we apply a whitening filter to $\boldsymbol{\varepsilon}_r$. Specifically,
considering that $\mathbf{C}_r$ is a positive-definite Hermitian matrix, the Cherosky decomposition is adopted as $\mathbf{C}_r^{-1}=\mathbf{U}_r\mathbf{U}_r^H$, where $\mathbf{U}_r$ is the lower triangle matrix. Then, $\mathbf{U}_r^H$ is used as the whitening filter in  \eqref{GLRT2},
\begin{equation}
\begin{cases}
 \mathcal{H}_0:\tilde{\mathbf{z}}_r =\mathbf{U}_r^H \boldsymbol{\varepsilon}_r ,\\
  \mathcal{H}_1:\tilde{\mathbf{z}}_r=\alpha_r \sqrt{L }\mathbf{U}_r^H \mathbf{d}(\varphi _r,\theta ) +\mathbf{U}_r^H\boldsymbol{\varepsilon}_r,
\end{cases}
\end{equation}
where $\mathbf{d}(\varphi _r,\theta )=\mathrm{vec} (\mathbf{B}_r\mathbf{W})$. Thus, the corresponding GLRT detector is given by
\begin{equation}\label{panjue2}
\Delta _r =\frac{\max _{\alpha_r,\varphi _r,\theta }f(\tilde{\mathbf{z} }_r|\alpha_r,\varphi _r,\theta,\mathcal{H} _1 ) }{f(\tilde{\mathbf{z} }_r|\mathcal{H} _0 )} \mathop{\gtrless}\limits_{\mathcal{H} _0}^{\mathcal{H} _1}\zeta ,
\end{equation}
where $f(\tilde{\mathbf{z} }_r|\alpha_r,\varphi _r,\theta,\mathcal{H} _1 )$ and $ f(\tilde{\mathbf{z} }_r|\mathcal{H} _0 )$  are the PDF under $\mathcal{H} _1$ and $\mathcal{H} _0$, respectively, and $\zeta$ is the decision threshold. For given $\varphi _r$ and $\theta$, the   maximum likelihood estimation (MLE) of $\alpha_r$ is obatined using the complex least-squares estimation and given as
\begin{equation}\label{alpha2}
    \hat{\alpha}_r=\frac{\mathbf{d}^{H}(\varphi _r,\theta) \mathbf{C}^{-1}_r \tilde{\mathbf{z}}_r}{\mathbf{d}^{H}(\varphi _r,\theta) \mathbf{C}^{-1}_r \mathbf{d}(\varphi _r,\theta)}.
\end{equation}
By substituting \eqref{alpha2} into \eqref{panjue2}, the MLE of $[\varphi _r,\theta]$ can be expressed as
\begin{equation}
[\hat{\varphi }_r,\hat{\theta}]=\arg \max _{\varphi _r,\theta} \frac{\big|\mathbf{d}^{H}(\varphi _r,\theta) \mathbf{C}^{-1}_r \tilde{\mathbf{z}}_r\big|^{2}}{\mathbf{d}^{H}(\varphi _r,\theta) \mathbf{C}^{-1}_r \mathbf{d}(\varphi _r,\theta)}.
\end{equation}
Hence, the GLRT test statistic is expressed as
\begin{align}
    \ln (\Lambda _r)&=\cfrac{\big|\mathbf{d}^{H}(\hat{\varphi }_r,\hat{\theta}) \mathbf{U}_r\mathbf{U}_r^{H} \tilde{\mathbf{z}}_r\big|^{2}}{\big\|\mathbf{U}^{H}_r \mathbf{d}(\hat{\varphi }_r,\hat{\theta})\big\|^{2}}\nonumber\\
    &=\cfrac{\big|\operatorname{tr}(\tilde{\mathbf{Z}}_r \mathbf{W }^H \mathbf{\hat{B}}^{H}_r \tilde{\mathbf{Q}}^{-1}_r)\big|^{2}}{\operatorname{tr}(\mathbf{\hat{B} }_r \hat{\mathbf{W}}\mathbf{\hat{B}}^{H}\tilde{\mathbf{Q}}^{-1}_r)}
    \mathop{\gtrless}\limits_{\mathcal{H} _0}^{\mathcal{H} _1}\ln (\zeta) ,
\end{align}
where
$
\tilde{\mathbf{Q}}_r=\mathbf{Q}_r+\sigma_{\rm n}^{2} \mathbf{I}_{N_1}.
$
The asymptotic distribution is expressed as
\begin{equation}
\ln(\Lambda_r ) \sim\left\{\begin{array}{l}
\mathcal{H}_{1}: \mathcal{X}_{2}^{2}(\rho_r), \\
\mathcal{H}_{0}: \mathcal{X}_{2}^{2},
\end{array}\right.
\end{equation}
where the non-centrality parameter $\rho_r$ of the $r$-th RAP is given as
\begin{align}
\rho_r &=\mathbb{E}\big(|\alpha_r|^{2} L \operatorname{vec}^{H}(\mathbf{B}_r\mathbf{W}) \mathbf{C}^{-1}_r \operatorname{vec}(\mathbf{B}_r\mathbf{W})\big) \nonumber\\
&=\sigma_{\rm rcs}^{2} L \operatorname{tr}\big(\mathbf{B}_r\hat{\mathbf{W}}\mathbf{B}_r^H(\mathbf{Q}_r+\sigma_{\rm n}^{2} \mathbf{I}_{N_1})^{-1}\big).
\end{align}
Besides, the non-centrality parameter $\rho_0$ of the BS is given as
\begin{align}
\rho_0 &=\mathbb{E}\big(|\alpha_0|^{2} L \operatorname{vec}^{H}(\mathbf{B}_0\mathbf{W}) \mathbf{C}^{-1}_0 \operatorname{vec}(\mathbf{B}_0\mathbf{W})\big) \nonumber\\
&=\sigma_{\rm rcs}^{2} L \operatorname{tr}\big(\mathbf{B}_0\hat{\mathbf{W}}\mathbf{B}_0^H(\sigma_{\rm n}^{2} \mathbf{I}_{N_1})^{-1}\big).
\end{align}
Similar to \eqref{p_d}, we can obtain the detection probability $P_{{\rm D}_r}$ of the $r$-th RAP  as well as $P_{{\rm D}_0}$ of the BS.

\subsection{Voting Aggregation}

The CC performs voting aggregation when receiving the binary inference results from the RAPs and BS, which can be modeled by
\begin{equation}
\begin{cases}
    \mathcal{H}_{0}: \text { No target, }\\
    \mathcal{H}_{1} \text { : Exist target. }
    \end{cases}
\end{equation}
Then the voting rule is expressed as
\begin{equation}
\begin{cases}
    \mathcal{H}_{0}: \sum_{r =0}^{R} D_{r} \leq \kappa ,\\
    \mathcal{H}_{1}: \sum_{r =0}^{R} D_{r} \geq \kappa,
    \end{cases}
\end{equation}
where $D_r$, $r\in\mathcal{R}'$ is the binary inference result, with $D_r = 0$ standing for no target and $D_r = 1$ standing
for an existing target, and $\kappa$ represents the voting threshold. The probability of error at the CC is \cite{1997Distributed}
\begin{align}\label{probability_of_error}
\Upsilon(&\kappa,\hat{P}_{\rm D})=\frac{1}{2}+ \frac{1}{2} \sum_{i=0}^{\kappa-1}\binom{R+1}{i}\cdot \nonumber \\
&\big[(\hat{P}_{\rm D})^{i}(1-\hat{P}_{\rm D})^{R+1-i}-(\hat{P}_{\rm FA})^{i}(1-\hat{P}_{\rm FA})^{R+1-i}\big],
\end{align}
where
\begin{equation}\label{p_d_sum}
\hat{P}_{\rm D}=\frac{1}{R+1} \sum_{r=0}^{R} P_{{\rm D}_r},
\end{equation}
and
\begin{equation}
\hat{P}_{\rm FA}=\frac{1}{R+1} \sum_{r=0}^{R} P_{{\rm FA}_r},
\end{equation}
$P_{{\rm D}_r}$ and $P_{{\rm FA}_r}$,$r\in\mathcal{R}'$ represent the detection probability and the false alarm probability, respectively, and
\begin{equation}
\binom{R+1}{i}=\cfrac{(R+1)!}{i !(R+1-i) !}.
\end{equation}
The optimal $\kappa$ is obtained as \cite{1997Distributed}
\begin{equation}\label{nt}
    \tilde{\kappa} =\min \left(R+1,\left \lceil \frac{R+1}{1+\beta(\hat{P}_{\rm D})} \right \rceil \right),
\end{equation}
where
\begin{equation}\label{beta}
\beta(\hat{P}_{\rm D})=\frac{\ln \frac{\hat{P}_{\rm FA}}{\hat{P}_{\rm D}}}{\ln \frac{1-\hat{P}_{\rm D}}{1-\hat{P}_{\rm FA}}}.
\end{equation}

\subsection{Problem Formulation}
We aim to minimize the probability of error  at the CC,  but the expression of $\Upsilon(\kappa,\hat{P}_{\rm D})$, as shown in \eqref{probability_of_error}, is quite complex.   To handle this issue, we first introduce the following lemmas.
\begin{lemma} \label{lemma1}
Given $\hat{P}_{\rm D}\in(0,1)$, $\beta(\hat{P}_{\rm D})$ decreases as $\hat{P}_{\rm D} $ increases.
\end{lemma}
\IEEEproof See Appendix A.

\begin{lemma} \label{lemma2}
Given $\hat{P}_{\rm D}\in(0,1)$, $\Upsilon(\tilde{\kappa}, \hat{P}_{\rm D})$ decreases as $\hat{P}_{\rm D} $ increases.
\end{lemma}

\IEEEproof  See Appendix B.

Based on \textbf{Lemma} \ref{lemma2},  the minimization of the probability of error at the CC is equivalent to the maximization of $\hat{P}_{\rm D}$. Thus, the optimization problem is formulated as
\begin{align*}
\mathcal{P}_3: \ &\max_{\mathbf{p\geq0}}\hat{P}_{\rm D} \\
\text {s.t.}\ &\gamma_{k} \geq \Gamma, k\in \mathcal{K},\\
& ||\mathbf{p}||_1 = P_{\max} ,
\end{align*}
which is a non-convex problem. Before giving the solution to problem $\mathcal{P}_3$, we first introduce the following lemma.
\begin{lemma}\label{lemma3}
When $p_0$ gradually increases and satisfies the transmit power constraint $||\mathbf{p}||_1 = P_{\max}$, $\hat{P_{\rm D}}$  gradually increases.
\end{lemma}
\IEEEproof See Appendix C.

\begin{figure}[htbp]
	\centerline{\includegraphics[width=0.51\textwidth]{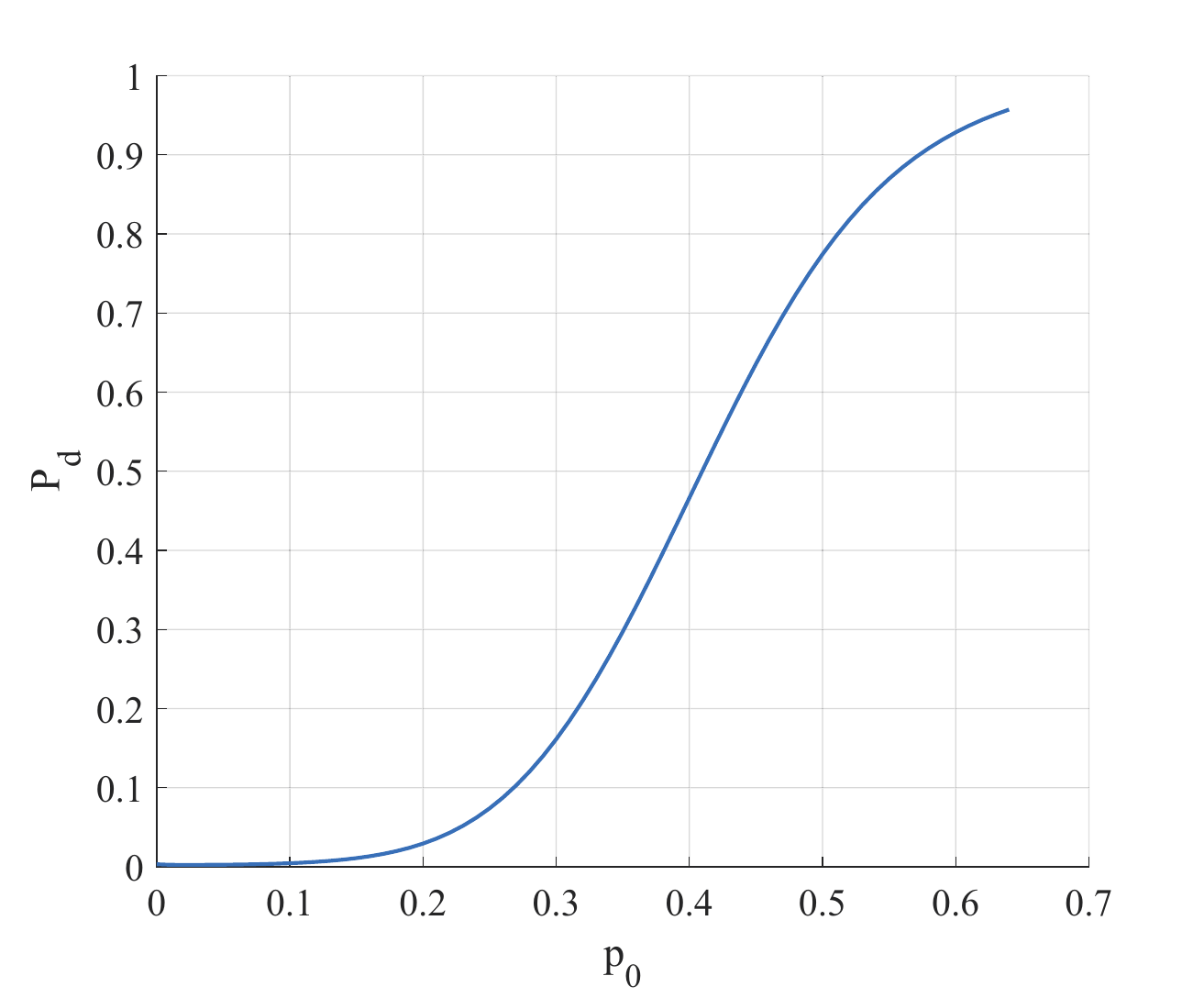}}
	\caption{ The relation between $\hat{P}_{\rm D}$ and $p_0$ with \{$P_{\max}=30$ dBm, $\Gamma=$15dB\}.}
	\label{p_d_and_p_0}
\end{figure}

For further validation of \textbf{Lemma} \ref{lemma3}, we plot the trend of  $\hat{P}_{\rm D}$ with respect to  $p_0$, as  shown in Fig. \ref{p_d_and_p_0}.  We first initialize $p_0=0$ and then  gradually increase $p_0$ with a step size of $\Delta p$. After obtaining the power allocation vector, $P_{{\rm D}_r}$ and $\hat{P}_{\rm D}$ are calculated using \eqref{p_d} and \eqref{p_d_sum}, respectively. The simulation result in Fig. \ref{p_d_and_p_0} is obtained by averaging over 1000 independent samples, which is consistent with the conclusion in \textbf{Lemma} \ref{lemma3}.

Based on \textbf{Lemma} \ref{lemma3}, we propose a heuristic algorithm and  summarize the proposed algorithm in Algorithm 1. Specifically, we decrease $p_0$ by $\Delta p$ gradually  and find the solution, $\mathbf{p}^{\prime}=[p_1,p_2,\ldots,p_K]^T$, to the power minimization problem under the SINR constraints. Such a process is repeated until $\|\mathbf{p}\|_1\leq P_{\max}$. In addition, the gap between the performance achieved by the heuristic algorithms and the optimal (upper bound) performance depends on  $\Delta p$. A smaller $\Delta p$ leads to better performance,  but also causes longer running  time. Fig. 3 shows the gap between the performance achieved by the heuristic algorithm and the optimal (upper bound) performance and Fig. 4 shows that the trade-off between the step size $\Delta_p$ and the time cost as well as the sensing performance, where the total non-centrality parameter is defined as $10\log_{10}(\sum_{r=0}^R \rho_r)$. Taking into account the time cost and sensing performance, we set $\Delta_p = 0.01$ in this work.

\begin{figure}[htbp]
	\centerline{\includegraphics[width=0.5\textwidth]{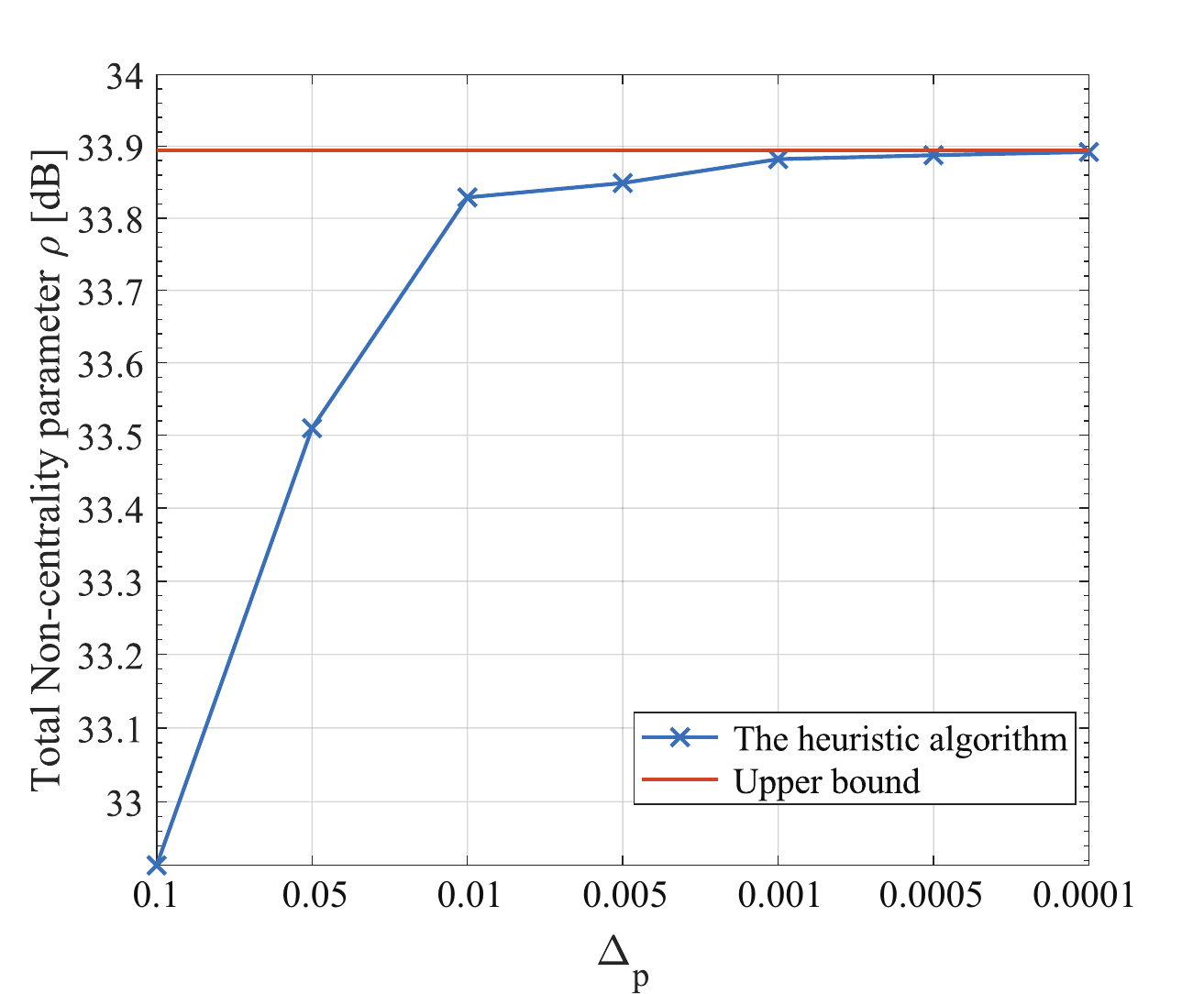}}
	\caption{ The performance gap between the heuristic algorithms and the upper bound.}
	\label{gap}
\end{figure}

\begin{figure}[htbp]
	\centerline{\includegraphics[width=0.5\textwidth]{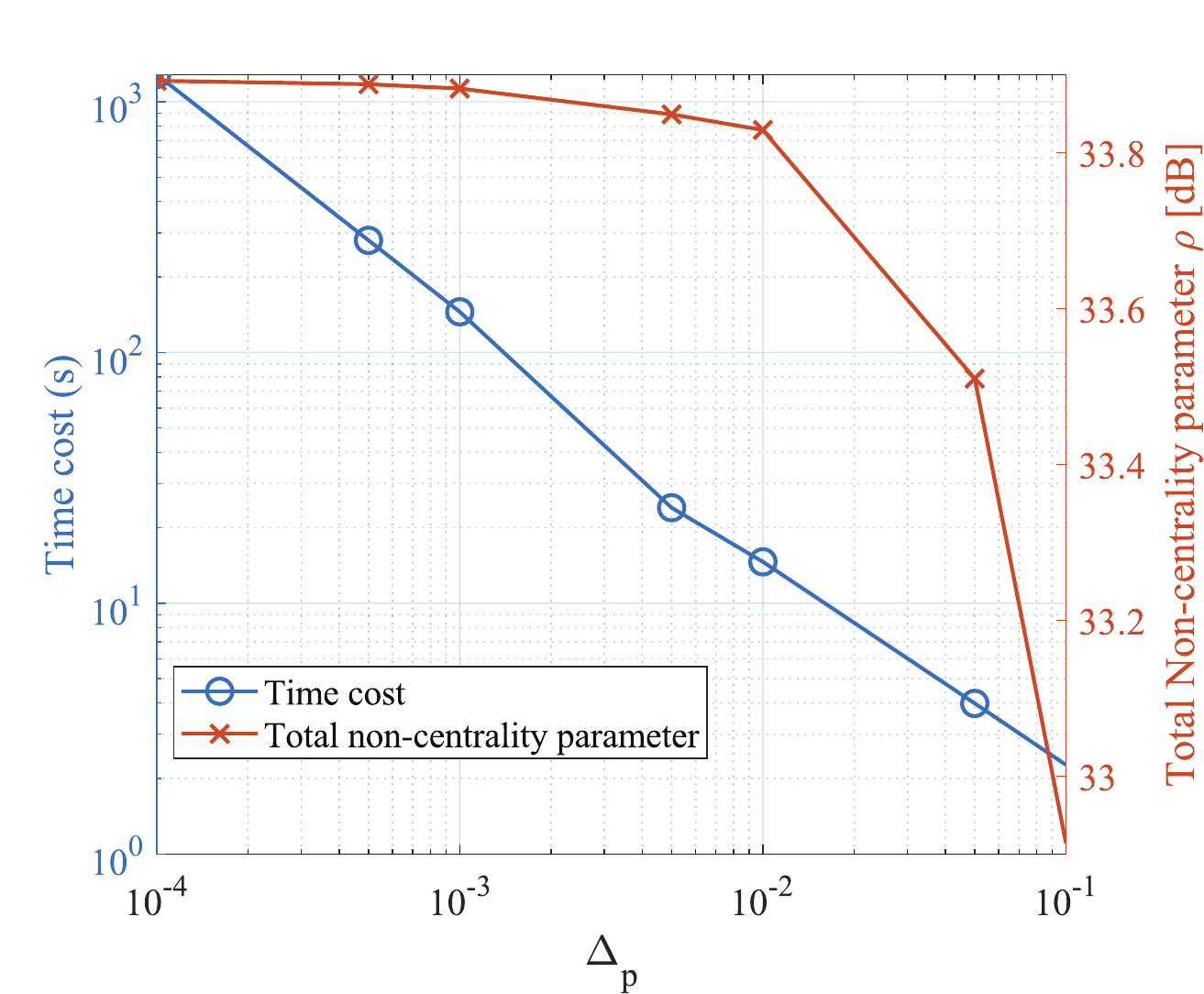}}
	\caption{ The trade-off between the step size $\Delta_p$ and the time cost as well as the sensing performance.}
	\label{Delta_p}
\end{figure}
\begin{algorithm}[htbp]
  \caption{Proposed heuristic algorithm of power allocation.}
  \label{alg::conjugateGradient}
  \begin{algorithmic}[1]
    \State Initialize $p_0'=P_{\max}$, the step size $\Delta p$, and $P_{\mathrm{sum}}=p_0'$.
    \While {$P_{\mathrm{sum}}>=P_{\max}$}
    \State Set $p_0=p_0'-\Delta p$.
    \State Find solution to  the following problem using SDP, i.e.,
    \begin{equation*}
    \mathcal{P}_{\rm{A}}:
    \begin{cases}
 \min_{\mathbf{p}^{\prime}}&\|\mathbf{p}\|_1\\
       \text {s.t.}  &\gamma_{k} \geq \Gamma,\quad k\in \mathcal{K}.
        \end{cases}
     \end{equation*}
    \State Set $P_{\mathrm{sum}}=\|\mathbf{p}\|_1$.
    \State Set $p_0'=p_0$.
    \EndWhile
    \State Output: $\mathbf{p}$.
  \end{algorithmic}
\end{algorithm}

\subsection{Complexity Analysis}
Note that the computational complexity of Algorithm 1 is mainly determined by finding solution to $\mathcal{P}_{\rm{A}}$ using SDP, which needs to be executed in each iteration. The corresponding SDP problem has $K$ nonnegative variables and $K$ linear inequality constraints. For a feasible instance of $\mathcal{P}_{\rm{A}}$, interior point methods can generate a $\mu$-optimal solution in $\mathcal{O} (\sqrt{K} \log (1 / \mu))$ iterations, each requiring at most $\mathcal{O} (K^3)$ arithmetic operations \cite{Karipidis2008QualityofService}. Then we set the complexity of solving $\mathcal{P}_{\rm{A}}$ as $\mathcal{O} (D)$. Hence, the complexity for Algorithm 1 is $\mathcal{O} (D_{\rm{iter}}D)$, where $D_{\rm{iter}}$ is the number of iterations and has the order of magnitude of $\mathcal{O} (\log(P_{\max}/\Delta p))$.

\section{Numerical Results}
\begin{figure}[htbp]
	\centerline{\includegraphics[width=0.49\textwidth]{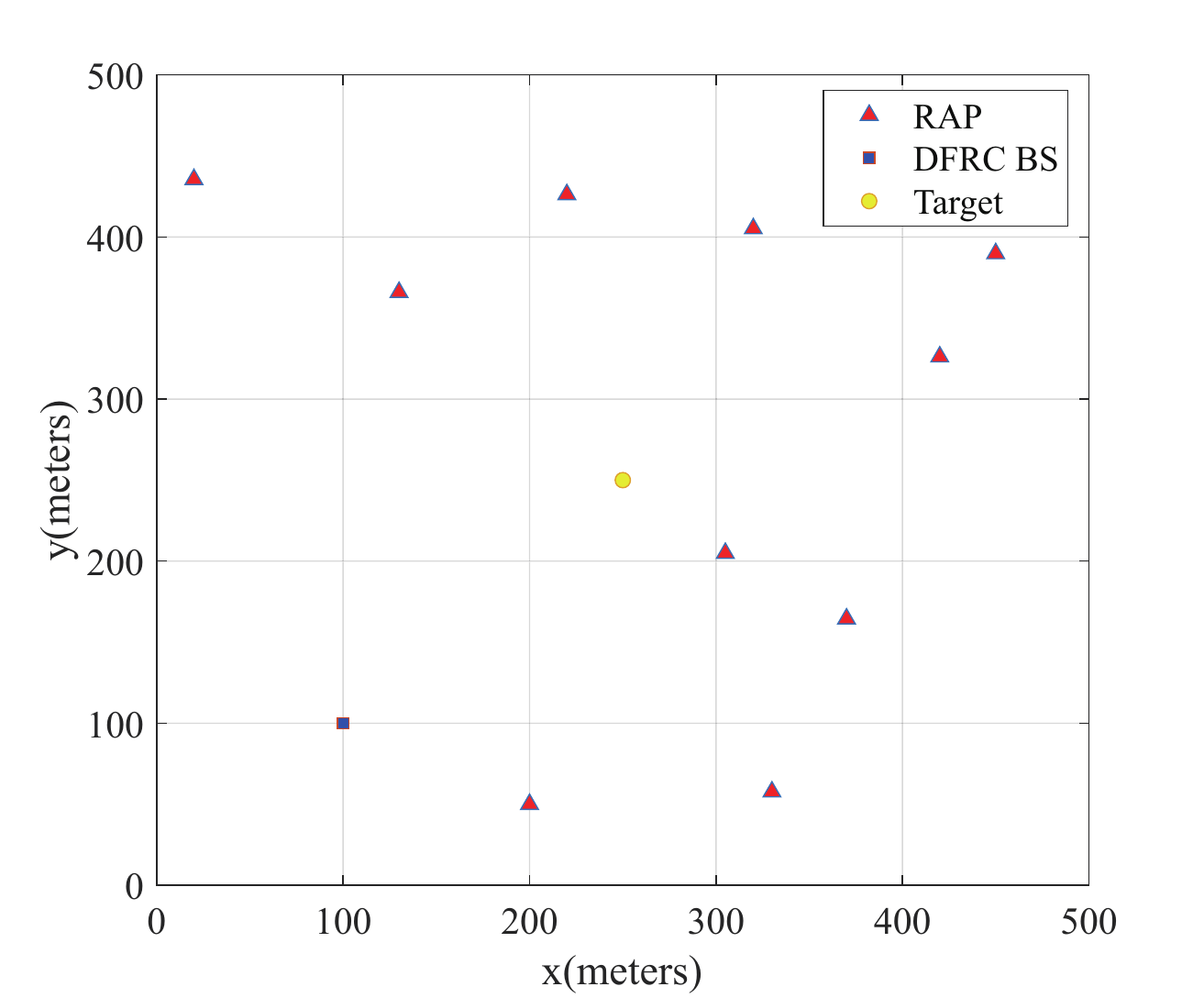}}
	\caption{  The 2D locations of the RAPs, DFRC BS, and the target.}
	\label{2D}
\end{figure}

To evaluate the performance of the proposed DFRC system, we perform numerical simulations in a 500 m $\times$ 500 m region with a BS, $R=10$ RAPs, and $K=8$ UEs  in the suburb area. The locations of the RAPs, UEs, and BS are randomly generated and the target is in the center of the region, as shown in Fig. \ref{2D}.
The BS and  RAPs each is equipped with $N_0=N_1$ = 20 receive antennas. Besides, the BS is equipped with $M$ = 16 transmit antennas.  Similar to  \cite{Xia2020DeepLearning}, the channel model is generated using $\mathbf{h}_{k}=\sqrt{m_{k}} \tilde{\mathbf{h}}_{k} \in \mathbb{C}^{M \times 1}$,
where $\tilde{\mathbf{h}}_{k} \sim \mathcal{C} \mathcal{N}(\mathbf{0}, \mathbf{I}_{M})$ is the small-scale fading and $m_{k}=128.1+37.6 \log _{10}(d)$ [dB] represents the path loss between the $k$-th UE and BS with $d$ being the distance in kilometers. The target-free channel $\mathbf{G}$ is generated using the same channel model.
The symbol number is set as $L$ = 30 and the detection threshold is determined by the false alarm probability $P_{\rm FA}$ = $10^{-5}$.

\begin{figure}[htbp]
	\centerline{\includegraphics[width=0.49\textwidth]{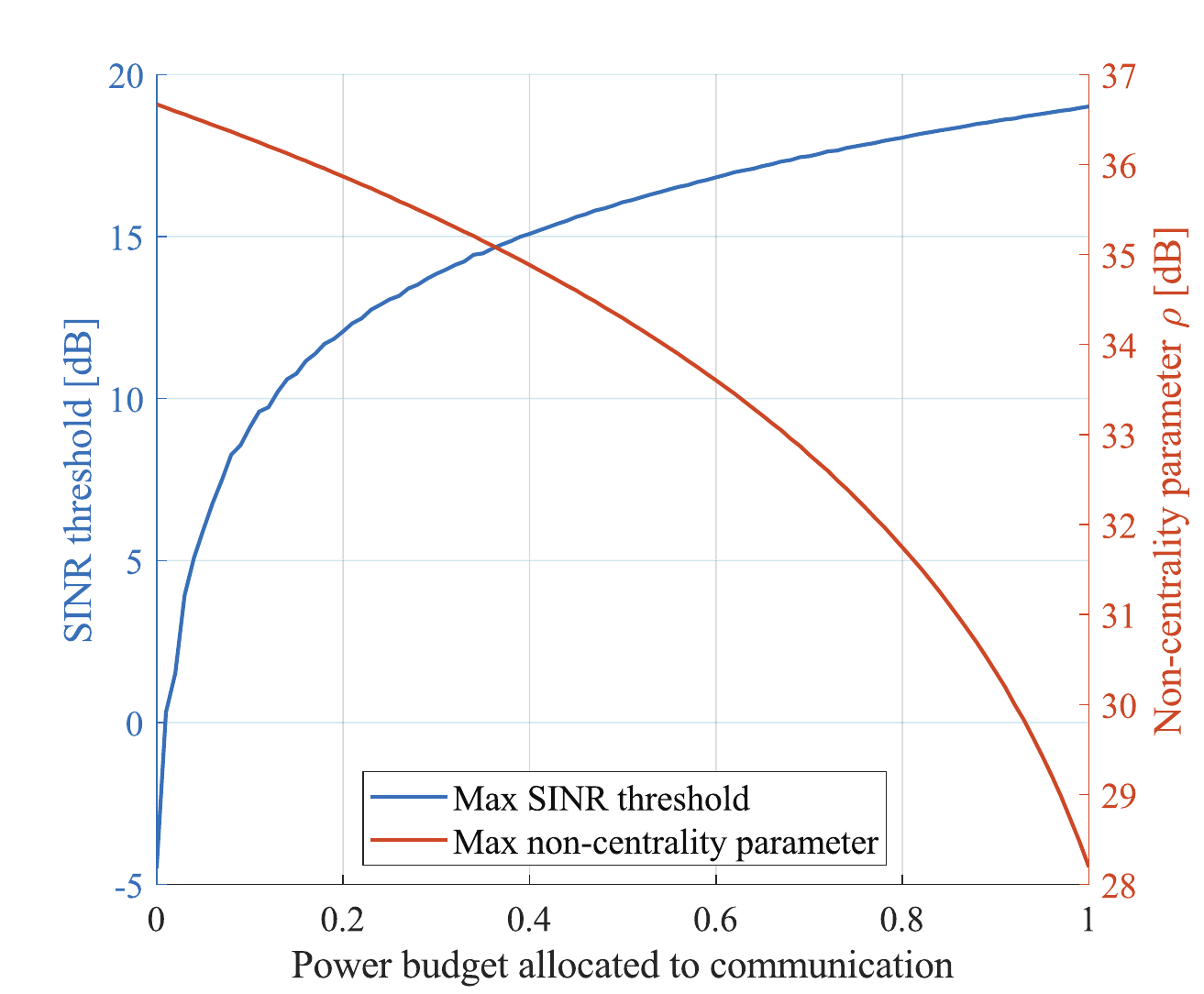}}
	\caption{  The trade-off between communication and sensing with respect to resource allocation.}
	\label{trade_off}
\end{figure}

As shown in Fig. \ref{trade_off}, the max SINR threshold is 0 when all power budget is allocated to sensing and the non-centrality parameter $\rho$ is the smallest when all power budget is allocated to communication. Meanwhile, it can be found that when SINR threshold is 15 dB, both communication and sensing can achieve acceptable performance. Therefore SINR threshold is  set as $\Gamma=$ 15 dB.

\subsection{Results with Unlimited Backhaul Capacity}
We first consider the case where the  backhaul capacity is unlimited. Before presenting the numerical results, we first introduce the following baseline schemes for comparison.
\begin{enumerate}
\item
  Firstly, we find the solution to problem $\mathcal{P}_2$ considering the cases with and without dedicated sensing symbols (marked as ``with $s_0$" and ``w/o $s_0$", respectively).
\item Then, only-active  and only-passive sensing schemes (marked as ``active" and ``passive", respectively) are introduced.
\item
  Finally, we also introduce the communication-centric scheme which aims to minimize the total power consumption (marked as ``$\min P_{\rm total}$'') under the same constraints for comparison.
\end{enumerate}

\begin{figure}[htbp]
	\centerline{\includegraphics[width=0.49\textwidth]{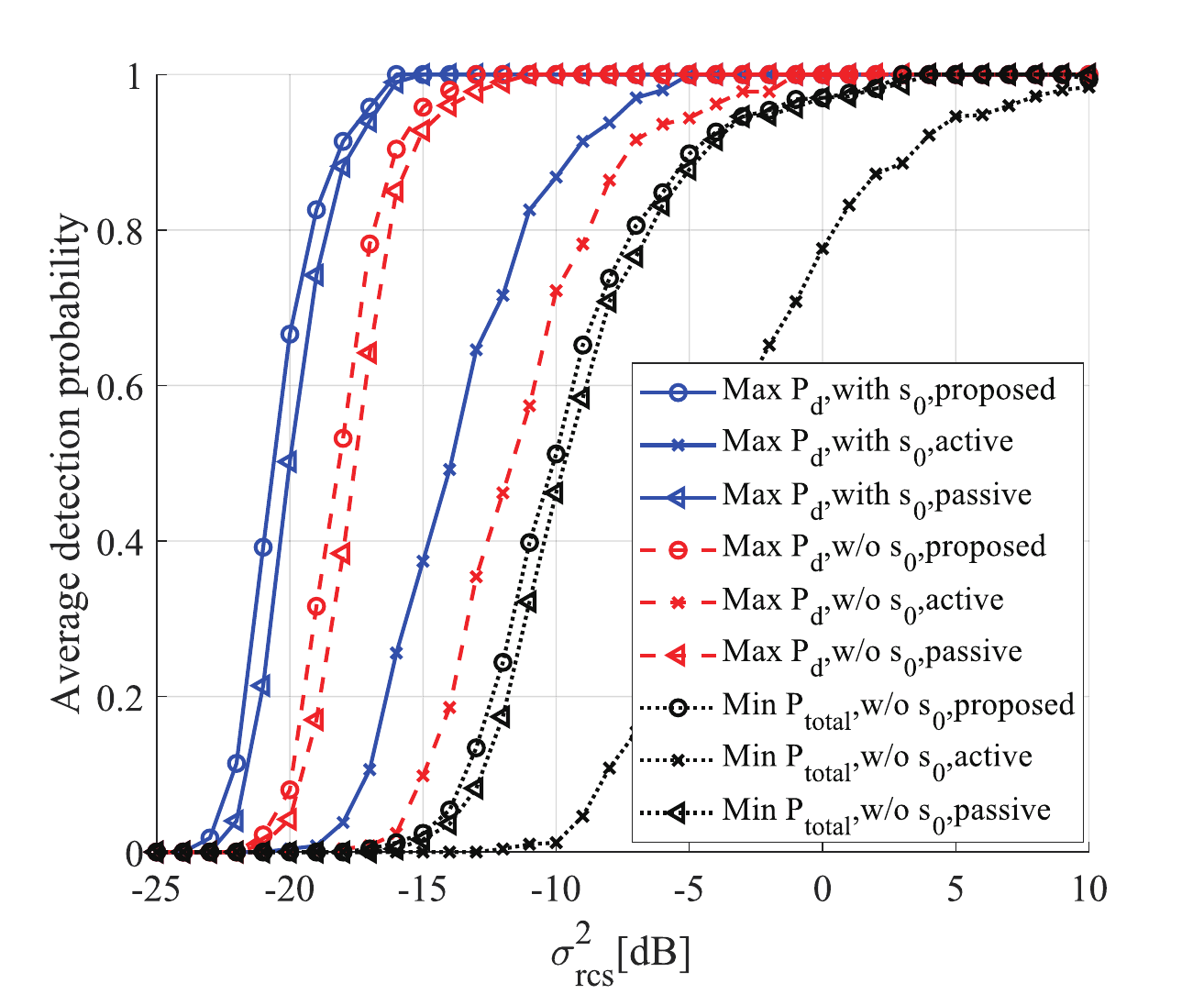}}
	\caption{ The average detection probability vs. $\sigma^2_{\rm rcs}$ with $P_{\max}=30$ dBm. }
	\label{fig1}
\end{figure}

We evaluate these schemes using average detection probability, which is calculated by averaging over 1000 random samples. Fig. \ref{fig1} shows that the average detection probability increases with the increase of $\sigma^2_{\rm rcs}$. We find the proposed IAPS scheme with sensing symbols has the best performance, whose average detection probability  can reach 1 when $\sigma_{\rm rcs}^{2} = -16$ dB. The performance of the only-passive sensing scheme is worse than that of the proposed IAPS scheme, but is much better than that of the only-active sensing scheme in both the cases with and without dedicated sensing symbols. This is because when the backhaul capacity is unlimited, all the RAPs can send the reflective signals they have received to the CC  and $R$ observed signals  can be utilized in  the only-passive sensing scheme. However, only the echo signal received at the BS is utilized in  the only-active  sensing scheme.  Besides, the average detection probability of the communication-centric scheme is almost 0 when $\sigma_{\rm rcs}^{2} \leq -16$ dB since the sensing requirement is not considered. We also observe that compared to the cases without  dedicated sensing symbols, the average detection probability of difference schemes  can be improved by using dedicated sensing symbols, which validates the importance of dedicated sensing symbols.\par

\begin{figure}[htbp]
	\centerline{\includegraphics[width=0.49\textwidth]{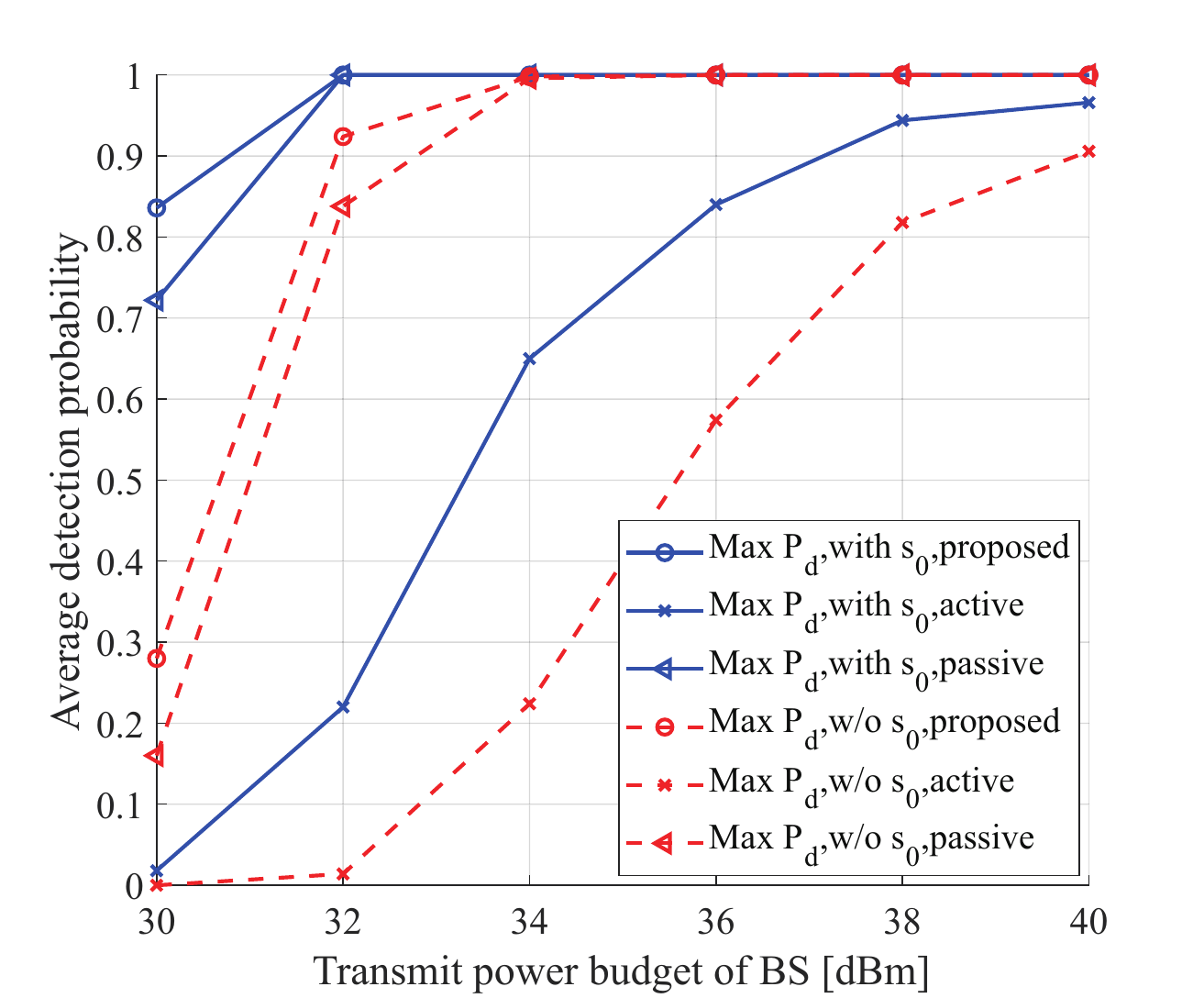}}
	\caption{ The average detection probability vs. the transmit power budget of the BS $P_{\max}$ with $\sigma_{\rm rcs}^{2}=-19$dB. }
	\label{fig2}
\end{figure}

Fig. \ref{fig2} shows  the average detection probability with respect to the transmit power budget of the BS $P_{\max}$. We find that the average detection probability can be improved  with more available power. Consistent  with  the results in Fig. \ref{fig1}, the proposed IAPS scheme shows better performance than the other two schemes  in both the cases with and without dedicated sensing symbols. Besides, with the advantage of quantity, the proposed scheme without dedicated sensing symbols, as well as the only-passive sensing scheme without dedicated sensing symbols, shows better performance than the only-active sensing scheme with dedicated sensing symbols. This is because the communication symbols can also be used for target detection. \par

\begin{figure}[htbp]
	\centerline{\includegraphics[width=0.49\textwidth]{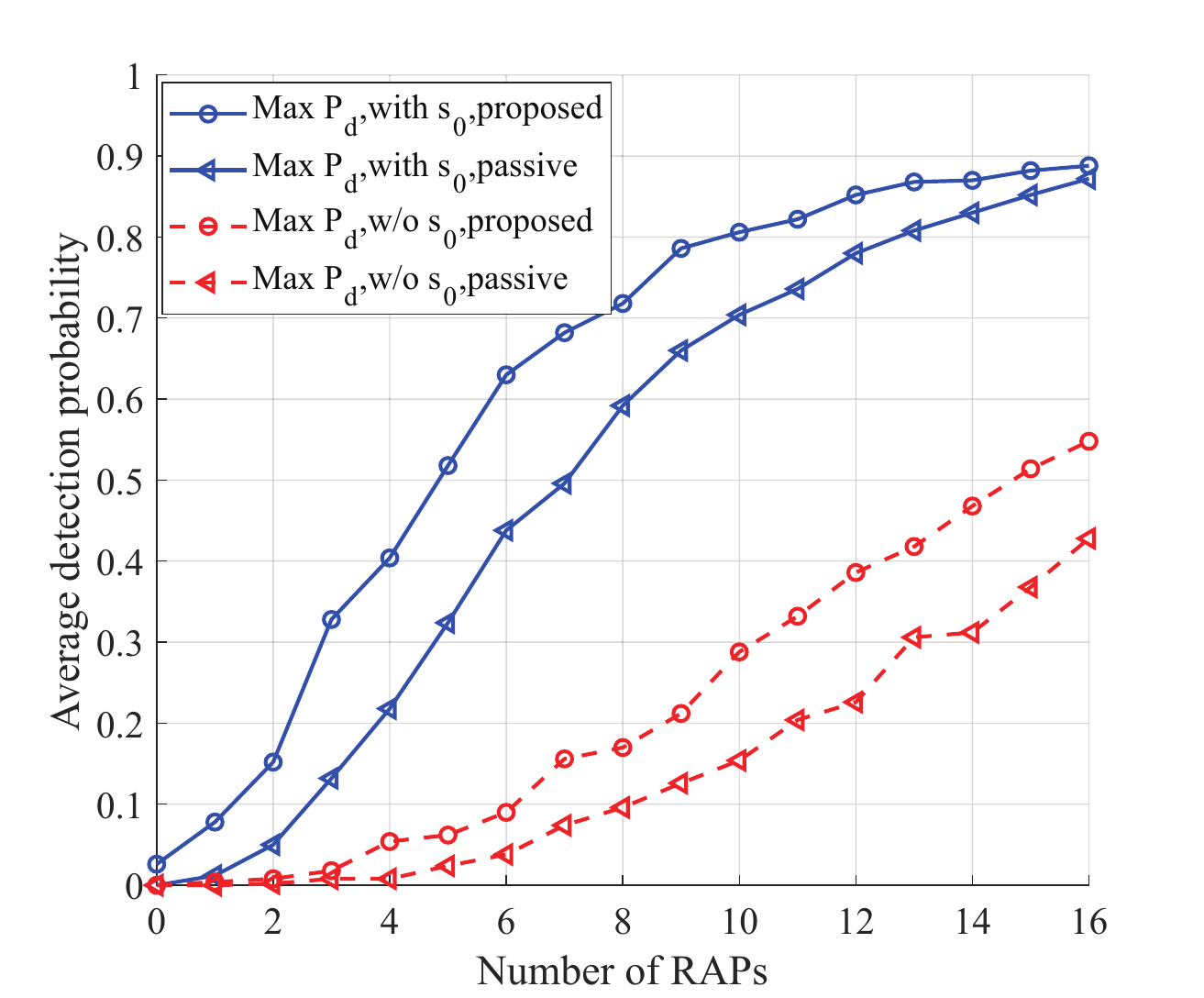}}
	\caption{ The average detection probability vs. the number of RAPs with \{$\sigma_{\rm rcs}^{2}=-19$dB, $P_{\max}=30$ dBm\}. }
	\label{fig4}
\end{figure}

Fig. \ref{fig4} further shows the relationship between the average detection probability and the  RAP number. The only-active sensing scheme is independent of the RAP number and thus is not presented in Fig. \ref{fig4}.  It is observed that   with the growth of the RAP number, the average detection probability of  the proposed IAPS scheme, as well as the only-passive sensing scheme, gradually increases. This fact indicates that the CC gains more information as the RAP number increases. In addition, we find that the use of the dedicated sensing symbols allows a fast convergence rate, compared to the scheme without dedicated sensing symbols. Finally, we observe that the proposed scheme is better than the only-passive sensing scheme and the only-passive sensing scheme requires $2-4$ more RAPs to achieve the same sensing performance as the proposed scheme.

\begin{figure}[htbp]
	\centerline{\includegraphics[width=0.49\textwidth]{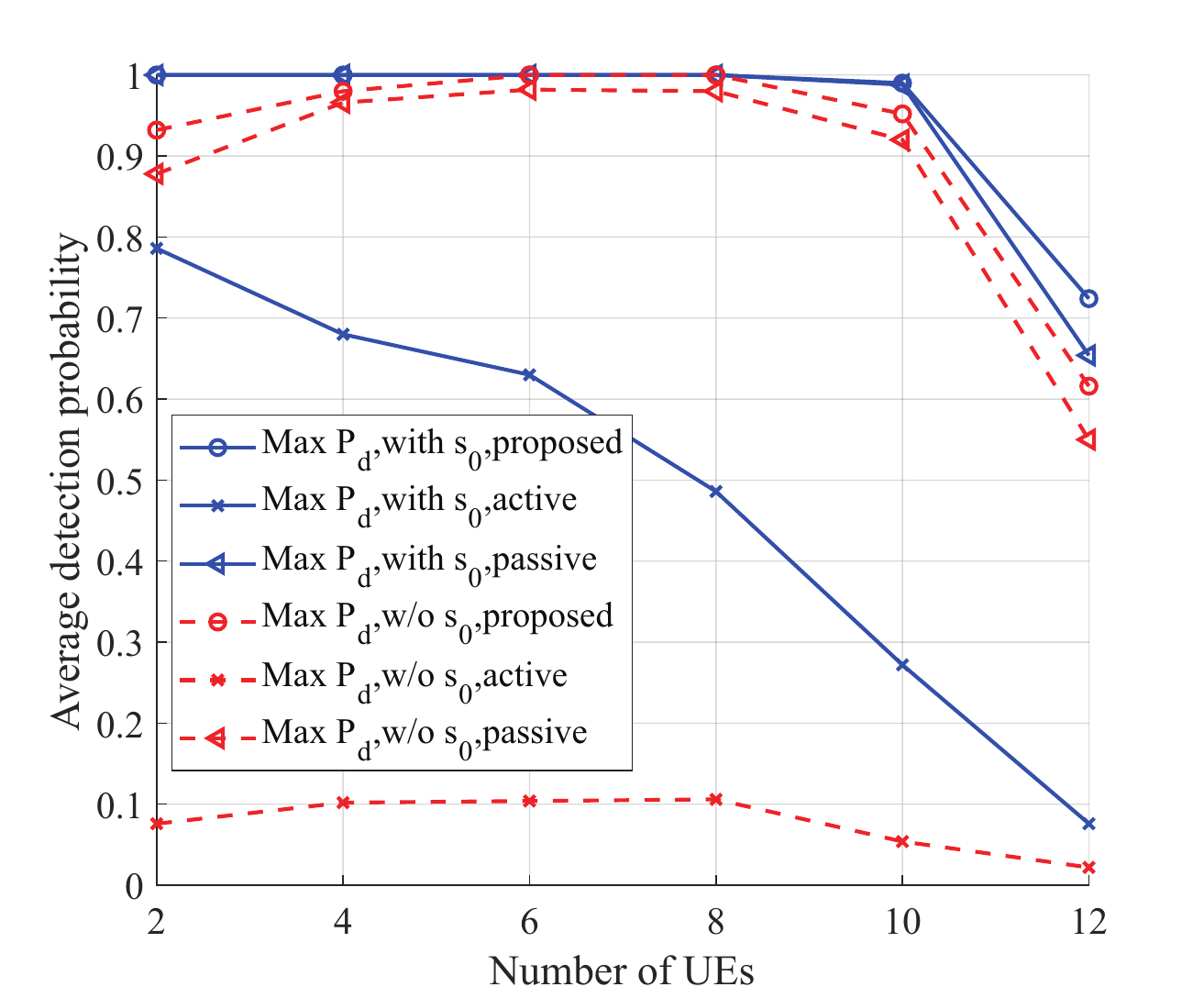}}
	\caption{ The average detection probability vs. the number of UEs with \{$\sigma_{\rm rcs}^{2}=-19$dB, $P_{\max}=33$ dBm\}. }
	\label{fig3}
\end{figure}

Fig. \ref{fig3} illustrates the effect of the number of UEs on the average detection probability. We observe that the average detection probability of the schemes with dedicated sensing symbols gradually decreases as the UE number increases. This is because, with the fixed power budget $P_{\max}$ and the increase of the UE number, more power is allocated to meet the user SINR constraints and less power is allocated to the sensing symbols. However, in the case without dedicated sensing symbols as the UE number increases, the average detection probability of the proposed IAPS scheme, as well as the only-passive sensing scheme, increases first and then decreases. The reason behind this phenomenon is that a small increase of the UE number leads to more communication symbols used for sensing and further the increase of the average detection probability. However, when the UE number exceeds a threshold value (i.e., 8 in Fig. \ref{fig3}),  more power is used to satisfy the increasing SINR constraints, which is not conducive to sensing performance.
\subsection{Results with Limited Backhaul Capacity}

\begin{figure}[htbp]
	\centerline{\includegraphics[width=0.49\textwidth]{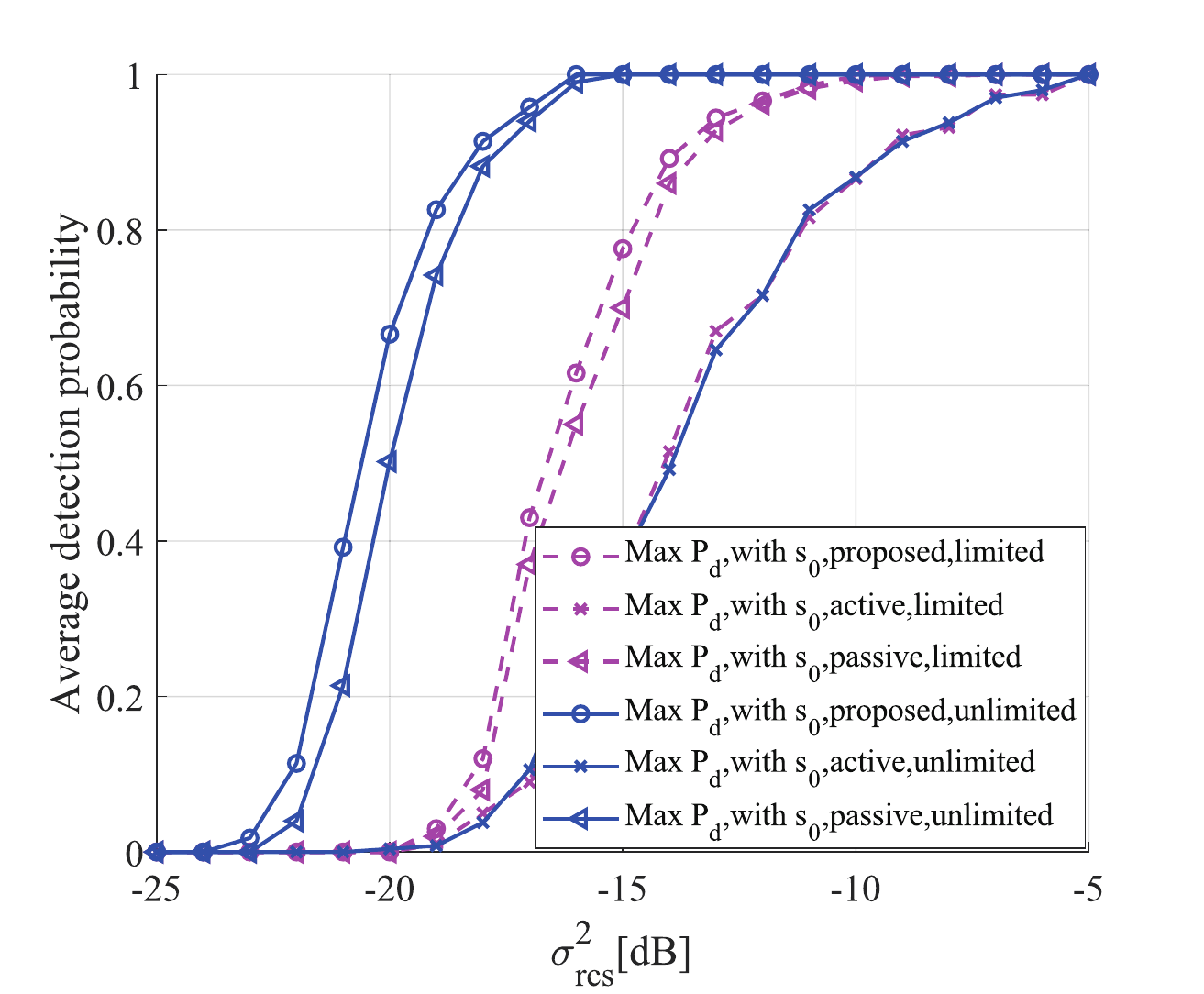}}
	\caption{ The average detection probability vs. $\sigma_{\rm rcs}^{2}$, $P_{\max}=30$ dBm. }
	\label{fig5}
\end{figure}

We first introduce two labels, i.e., ``unlimited" and ``limited" to distinguish the two cases with unlimit and limited backhaul capacity. Fig. \ref{fig5} shows that the average detection probability increases  with the increase of $\sigma_{\rm rcs}^{2}$, regardless of whether  backhaul capacity is unlimit or limited. We also observe that the achieved performance with the limited backhaul capacity  is worse than that  with the unlimited backhaul capacity because more sensing information is exploited in the case of unlimited backhaul capacity but only binary inference results are integrated in the  case of limited backhaul capacity. Similar to the results in Fig. \ref{fig1}, it is observed that the proposed IAPS scheme always achieves a higher average detection probability than that of the only-active and only-passive sensing schemes. Note that in the only-active sensing  scheme, the influence of backhaul capacity can be ignored because no passive sensing is involved.\par

\begin{figure}[htbp]
	\centerline{\includegraphics[width=0.49\textwidth]{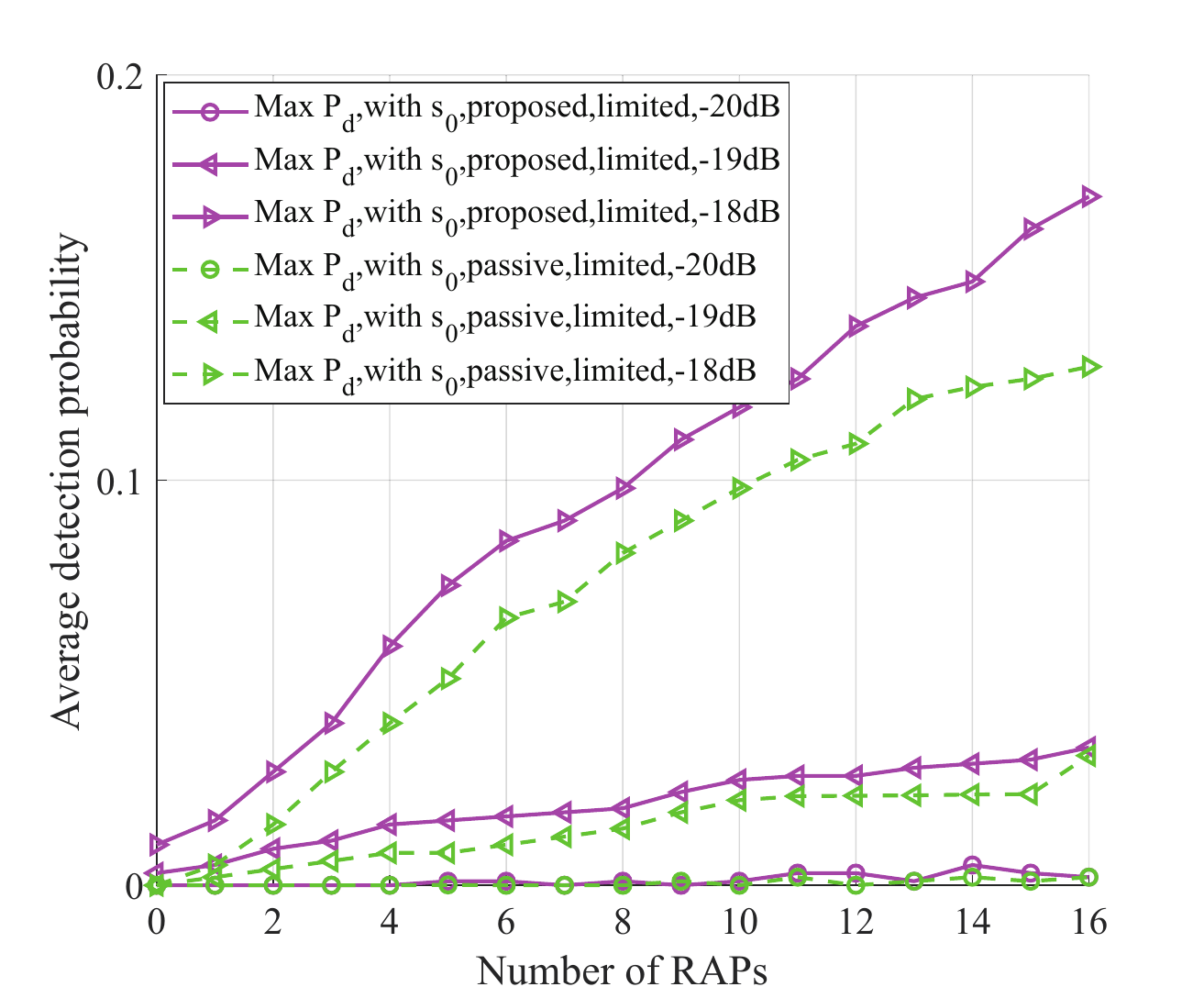}}
	\caption{ The average detection probability vs. the number of RAPs, $P_{\max}=30$ dBm.}
	\label{fig6}
\end{figure}

Fig. \ref{fig6} shows the average detection probability with respect to the number of RAPs with different $\sigma_{\rm rcs}^{2}$ values. The increase of the RAP number can reduce the probability of misjudgment after the voting aggregation. We note that the average detection probability increases as the number of the RAPs increases when $\sigma_{\rm rcs}^{2}\geq-19$ dB,  but the curve with $\sigma_{\rm rcs}^{2}= -18$dB rises faster than  the curve with $\sigma_{\rm rcs}^{2}=-19$dB. In addition, we also find that when $\sigma_{\rm rcs}^{2}=-20$dB,  the average detection probability is almost unchanged. This is because the voting aggregation at the CC highly depends on the binary inference results of each single RAP and the detection probability of each single RAP relies on $\sigma_{\rm rcs}^{2}$.

\section{CONCLUSION}
In this paper, we proposed power optimization for IAPS and multi-user communications in DFRC systems, for the cases of the unlimited and limited backhaul capacity between RAPs and CC, respectively. In particular, we adopted different fusion schemes to exploit the active sensing and passive sensing information for different cases. On the one hand, the BS and RAPs sent  the sensing signal they have received to the CC for signal fusion when the backhaul capacity is unlimited. On the other hand, the RAPs made decisions and sent binary inference results to the CC for result fusion via voting aggregation when the backhaul capacity is limited. We then formulated optimization problems to maximize the detection probability under communication QoS constraints. Especially, we proposed a heuristic algorithm to solve complex non-convex problem in the cases of the limited backhaul capacity. Finally, numerical simulations in the cases with unlimited and limited backhaul capacity are conducted, which validate the performance gain of the proposed IAPS scheme and the positive effect of dedicated sensing symbols. Besides, it also reveals the effectiveness of the proposed fusion schemes. \par

\section*{Appendix}
\subsection{Proof of \textbf{Lemma} 1}
The derivative of $\beta(\hat{P}_{\rm D})$ is given as
\begin{align}\label{beta_derivative}
&\frac{{\rm d} \beta(\hat{P}_{\mathrm{D}})}{{\rm d} \hat{P}_{\mathrm{D} }}=\nonumber \\
&\frac{(\!1\!-\!\hat{P}_{\mathrm{D}})\big(\!\ln(\!1\!-\!\hat{P}_{\mathrm{FA}})\!-\!\ln(\!1\!-\!\hat{P}_{\mathrm{D}}\!)\!\big)\!+\!\hat{P}_{\mathrm{D}}
(\!\ln\hat{P}_
{\mathrm{FA}}\!-\!\ln\hat{P}_{\mathrm{D}}\!)}
{\hat{P}_{\mathrm{D}}(\!1-\hat{P}_{\mathrm{D}}\!)\big(\!\ln(\!1\!-\!\hat{P}_{\mathrm{D}}\!)\!-\!\ln(\!1\!-\!\hat{P}_{\mathrm{FA}}\!)\big)^2},
\end{align}
whose  denominator  is greater than 0. Define $\beta_1(\hat{P}_{\mathrm{D}})=(1-\hat{P}_{\mathrm{D}})\big(\ln(1-\hat{P}_{\mathrm{FA}})-\ln(1-\hat{P}_{\mathrm{D}})\big)+
\hat{P}_{\mathrm{D}}(\ln\hat{P}_{\mathrm{FA}}-
\ln\hat{P}_{\mathrm{D}})$. The first and second derivatives of $\beta_1(\hat{P}_{\rm D})$ are given as
\begin{align}\label{beta1_derivative}
\frac{{\rm d} \beta_1(\hat{P}_{\mathrm{D}})}{{\rm d} \hat{P}_{\mathrm{D} }}
\!=\!\ln(1\!-\!\hat{P}_{\mathrm{D}})\!-\!\ln(1\!-\!\hat{P}_{\mathrm{FA}})\!+\!\ln\hat{P}_{\mathrm{FA}}\!-\!\ln\hat{P}_{\mathrm{D}},
\end{align}
and
\begin{align}\label{beta1_2derivative}
\frac{{\rm d}^2 \beta_1(\hat{P}_{\mathrm{D}})}{{\rm d} (\hat{P}_{\mathrm{D} })^2}
=-\frac{1}{1-\hat{P}_{\mathrm{D}}}-\frac{1}{\hat{P}_{\mathrm{D}}},
\end{align}
respectively.
Note that \eqref{beta1_2derivative} is less than 0 because of $\hat{P}_{\rm D}\in(0,1)$, which suggests that  $\beta_1(\hat{P}_{\mathrm{D}})$ is a concave function and its maximum (i.e., 0) is achieved at the point  $\hat{P}_{\mathrm{D}}=\hat{P}_{\mathrm{FA}}$. We further have $\beta_1(\hat{P}_{\mathrm{D}})\leq0$  and $\frac{{\rm d} \beta(\hat{P}_{\mathrm{D}})}{{\rm d} \hat{P}_{\mathrm{D} }}\leq0$. Thus, $\beta(\hat{P}_{\rm D})$ decreases as $\hat{P}_{\rm D} $ increases. \hfill {$\blacksquare $}

\subsection{Proof of \textbf{Lemma} 2}
Based on \textbf{Lemma} \ref{lemma1}, $\left \lceil \frac{R+1}{1+\beta(\hat{P}_{\rm D})} \right \rceil$ follows a stepwise ascent as $\hat{P}_{\rm D} $ increases.  We first define  $\hat{P}^{\tilde{\kappa},\min}_{\rm D}$ and $\hat{P}^{\tilde{\kappa},\max}_{\rm D}$, which satisfy
\begin{equation}\label{lb_pd}
\hat{P}^{\tilde{\kappa},\min}_{\rm D}\!=\!\arg\min_{\hat{P}_{\rm D}}\Bigg[\!\min \Bigg(\!R+1,\left \lceil \frac{R+1}{1+\beta(\hat{P}_{\rm D})} \right \rceil \!\Bigg)=\tilde{\kappa}\!\Bigg],
\end{equation}
and
\begin{equation}\label{up_pd}
\hat{P}^{\tilde{\kappa},\max}_{\rm D}\!=\!\arg\max_{\hat{P}_{\rm D}}\Bigg[\!\min \Bigg(\!R+1,\left \lceil \frac{R+1}{1+\beta(\hat{P}_{\rm D})} \right \rceil \!\Bigg)=\tilde{\kappa}\!\Bigg],
\end{equation}
respectively.  $\kappa$ is a constant and equal to $\tilde{\kappa}$ when $\hat{P}_{\rm D}\in[\hat{P}^{\tilde{\kappa,\min}}_{\rm D},\hat{P}^{\tilde{\kappa,\max}}_{\rm D}]$.
We will prove \textbf{Lemma} \ref{lemma2} in two steps. Specifically, we first prove that the probability of error $\Upsilon\big(\kappa,\hat{P}_{\rm D}\big)$  at the CC decreases as $\hat{P}_{\rm D}$ increases when $\hat{P}^{\tilde{\kappa}}_{\rm D}\in(\hat{P}^{\tilde{\kappa}}_{\rm D_{\min}},\hat{P}^{\tilde{\kappa}}_{\rm D_{\max}}]$.
In such a case, $\tilde{\kappa}$ is fixed and equal to $\tilde{\kappa}$ and $\Upsilon(\kappa,\hat{P}_{\rm D})$ can be simplified as
  \begin{align}
  \Upsilon_1(\hat{P}_{\rm D})=\frac{1}{2} \sum_{i=0}^{\tilde{\kappa}-1}\binom{R+1}{i}(\hat{P}_{\rm D})^{i}(1-\hat{P}_{\rm D})^{R+1-i},
  \end{align}
 which decreases as $\hat{P}_{\rm D}$ increases. This is because the CDF of the binomial distribution $\Upsilon_1(\hat{P}_{\rm D})$ can  be represented in terms of the regularized incomplete beta function \cite{wadsworth1961introduction}:
  \begin{equation*}
  \Upsilon(\hat{P}_{\rm D})
  =(R+2-\tilde{\kappa})\binom{R+1}{i} \int_{0}^{1-\hat{P}_{\mathrm{D} }} t^{R+1-\tilde{\kappa}}(1-t)^{\tilde{\kappa}-1} {\rm d}t,
  \end{equation*}
  whose derivative meets the following constraint,
  \begin{align}
  &\frac{{\rm d} \Upsilon(\hat{P}_{\rm D})}{{\rm d} \hat{P}_{\mathrm{D} }}=\nonumber\\&
  -(R+2-\tilde{\kappa})\binom{R+1}{i} (1-\hat{P}_{\mathrm{D} })^{R+1-\tilde{\kappa}}(\hat{P}_{\mathrm{D} })^{\tilde{\kappa}-1}\leq0.
  \end{align}

Secondly, we prove that the probability of error $\Upsilon(\kappa,\hat{P}_{\rm D})$  at the CC  decreases when  $\kappa$ varies from $\tilde{\kappa}$ to $\tilde{\kappa}+1$ due to the increase of $\hat{P}_{\rm D}$. According to \eqref{lb_pd} and \eqref{up_pd}, we have
  \begin{align}\label{k+1-k}
     &\Upsilon(\tilde{\kappa}+1,\hat{P}^{\tilde{\kappa}+1,\min}_{\rm D})-  \Upsilon(\tilde{\kappa},\hat{P}^{\tilde{\kappa},\max}_{\rm D})\nonumber\\
     \leq&\Upsilon(\tilde{\kappa}+1,\hat{P}^{\tilde{\kappa},\max}_{\rm D})-\Upsilon(\tilde{\kappa},\hat{P}^{\tilde{\kappa},\max}_{\rm D})\nonumber\\
   =&\binom{R+1}{\tilde{\kappa}}\big[(\hat{P}^{\tilde{\kappa},\max}_{\rm D_{}})^{\tilde{\kappa}}(1-\hat{P}^{\tilde{\kappa},\max}_{\rm D})^{R+1-\tilde{\kappa}}\nonumber\\
   &-(\hat{P}_{\rm FA})^{\tilde{\kappa}}(1-\hat{P}_{\rm FA})^{R+1-\tilde{\kappa}}\big],
  \end{align}
and
  \begin{align}\label{k=}
   \tilde{\kappa}=\frac{R+1}{1+\beta(\hat{P}^{\tilde{\kappa},\max}_{\rm D})} = \frac{(R+1)\ln {\frac{1-\hat{P}_{\rm FA}} {1-\hat{P}^{\tilde{\kappa},{\max}}_{\rm D}}}}{\ln \frac{\hat{P}^{\tilde{\kappa},{\max}}_{\rm D}(1-\hat{P}_{\rm FA})} {\hat{P}_{\rm FA}(1-\hat{P}^{\tilde{\kappa},\max}_{\rm D})}}.
  \end{align}
 Then, substituting \eqref{k=} into \eqref{k+1-k}, we obtain
  \begin{align}
    \Upsilon(\tilde{\kappa}+1,\hat{P}^{\tilde{\kappa},{\max}}_{\rm D})- \Upsilon(\tilde{\kappa},\hat{P}^{\tilde{\kappa},{\max}}_{\rm D})=0
  \end{align}
which suggests that $\Upsilon(\kappa, \hat{P}_{\rm D})$ decreases when $\kappa$ varies from $\tilde{\kappa}$ to $\tilde{\kappa}+1$.
Finally, we can conclude that $\Upsilon(\kappa, \hat{P}_{\rm D})$ decreases as $\hat{P}_{\rm D} $ increases in the whole feasible region $\hat{P}_{\rm D}\in(0,1)$.\hfill {$\blacksquare $}

\subsection{Proof of \textbf{Lemma} 3}

Since $\hat{P}_{\rm D}$ is the average of $P_{{\rm D}_r}$ and $P_{{\rm D}_r}$ is a monotonically increasing function with respect to $\rho_r$ \cite{1993Fundamentals}, we just need to prove that $\rho_r$ increases as $p_0$ increases.
It is observed that $ \operatorname{tr}(\mathbf{Q}_{r}+\sigma_{\rm n}^{2} \mathbf{I}_{N_1}) $ is a constant associated with $P_{\max}$, and we also have
\begin{align}
\operatorname{tr}(\mathbf{B}_{r} \hat{\mathbf{W}}\mathbf{B}_{r}^{H})
&=\operatorname{tr}\big(\hat{\mathbf{W}}\mathbf{a}(\theta )\mathbf{b}_1^H(\varphi _r)\mathbf{b}_1(\varphi _r)\mathbf{a}^H(\theta ) \big)\nonumber\\
&\simeq \operatorname{tr}\big(\hat{\mathbf{W}} \mathbf{a}(\theta )\mathbf{a}^H(\theta ) \big).
\end{align}
where $\simeq$ represents  the same trend with respect to the variation of $p_0$ on the left-hand and right-hand sides.

Then,
\begin{align}
\rho_r &\simeq\operatorname{tr}\bigg (\mathbf{B}_{r} \hat{\mathbf{W}} \mathbf{B}_{r}^{H}\bigg ) \nonumber\\
&= \operatorname{tr}\big(\operatorname{diag}(\mathbf{p}) \tilde{\mathbf{W}}^{H}\mathbf{a}(\theta )\mathbf{a}^H(\theta )\tilde{\mathbf{W}} \big).
\end{align}
Obviously, $\mathbf{\tilde{w}}_0^H\mathbf{a}(\theta )> \mathbf{\tilde{w}}_i^H\mathbf{a}(\theta )$ due to the ZFR precoder in \eqref{w0}. Thus, $\rho_r$ increases as $p_0$ increases.\hfill {$\blacksquare $}\par

\bibliographystyle{IEEEtran}
\bibliography{reference}

\end{document}